\documentclass[journal,11pt]{IEEEtran}
\usepackage{graphicx}

\usepackage{epsfig}
\usepackage{times}
\usepackage[nocompress]{cite}
\usepackage{amsmath,amssymb}
\usepackage{pifont}
\usepackage{balance,simplemargins}
\usepackage{graphicx}
\usepackage{epsf}
\usepackage{xcolor}
\usepackage{simplemargins}

\begin{document}
\title{Analytical Modeling and Design of  Fresnel Lens Transducers
\author{Kapil~Dev$^\dagger$, Vibhu~Vivek$^\ddagger$, Babur~Hadimioglu$^\ddagger$ and Yehia~Massoud$^\dagger$\\
$^\dagger$Department of Electrical and Computer Engineering, Rice University, Houston TX 77005\\
$^\ddagger$Microsonic Systems San Jose, CA 95134; vibhu.vivek@microsonics.com; massoud@rice.edu}}
\maketitle

\footnotetext{This manuscript is an extension of an earlier work~\cite{kdevICES2011-1}.}

\begin{abstract}
In this paper, we present an analytical modeling technique for circularly symmetric piezoelectric transducers, also called as  Fresnel Lens. We also present the design of a flat/piston transducer that can generate unique acoustic wave patterns, having both converging and vortexing effects. The converging effect is generated by designing the transducer electrodes in the shapes of circular rings using Fresnel formula and exciting it with an RF signal of resonant frequency. The vortexing effect is achieved by cutting the rings to different sector angles: $90^{\circ}$, $120^{\circ}$,  $180^{\circ}$ and $270^{\circ}$. We use the analytical model to simulate the performance of these transducers. 
\end{abstract}

\vspace{-0.2in}
\section{Introduction}
\label{sec:intro}
Ultrasonics based microfluid processing systems are gaining popularity in biomedical, pharmaceutical and drug discovery applications because they provide faster, controlled and non-contact fluid-processing ability at low volumes (nL to hundreds of $\mu$L)~\cite{Yakub2005, HuangOSU2010}. Most of the ultrasonic systems today use lead zirconate titanate ($Pb[Zr_xTi_{1-x}]O_3, 0<x<1$) as piezoelectric material because it has higher electromehanical conversion coefficient as compared to other materials. It is also called PZT and there are different variations of PZT (PZT4, PZT5A, PZT5H etc.) depending on the relative composition of different elements and the temperature and pressure under which the ceramic is fabricated. In~\cite{SpicciCOMSOL2010,GutierrezCCE2010,GutierrezLIMA2010}, the piezoelectric behavior is modeled using finite-element method. In transmit and receive applications based on ultrasound today, the metal-electrodes are plated on flat top and bottom surfaces of the transducer and the transducer is excited in its thickness mode. Such a transducer with complete top and bottom surface as electrode have fixed "natural" focal-length, that is at the boundary of near-field and far-field regions, and it requires the sample or object under investigation to be placed at a fixed distance from the transducer~\cite{GSKinoBook1987}. For example, if a circular transducer has diameter of 8mm and the operating frequency is 4MHz, the natural focus of such a transducer in water (sound speed 1480m/s) is fixed at $\frac{D^2*frequency}{4*sound\_speed}$ = 43.2 mm. The focal length in this case is dependent only on the diameter of the electrode (the electrodes are kept circular shape due to their circularly symmetric acoustic field). To decrease the focal length, we need to reduce the size of transducer, which is not always desirable because smaller size transducer has lesser absolute sound intensity than a large size transducer.

In this paper, we present an analytical model for piezoelectric tFresnel Lens ransducers using the acoustic potential~\cite{kdevICES2011-1}. We design a transducer with an electrode having a concentric ring-shape. Depending on the ring-radii of different rings, such a transducer can be designed to have any focal length. As expected, such a transducer could provide optimum/maximum vertical component of particle displacement, but it provides negligible lateral rotational component, required for fluid processing applications~\cite{Xzhu98, HuangMEMS2001}. To increase the rotational component, we modify the electrode shape to sectored rings~\cite{MSSpatent2012, MSSpatent2012_EP, MSSpatent2012_JP, MSSpatent2012_WO, kdevICES2011-2, kdevMSthesis2011}. We analyze different sector angles of the transducer and conclude that $90^{\circ}$ sectored transducer gives maximum rotational component per unit electrode area; hence, it is the best candidate for the micro fluidic applications.

\begin{figure*}[t]
\begin{center}
\begin{tabular}{cc}
\includegraphics[width=0.9\columnwidth]{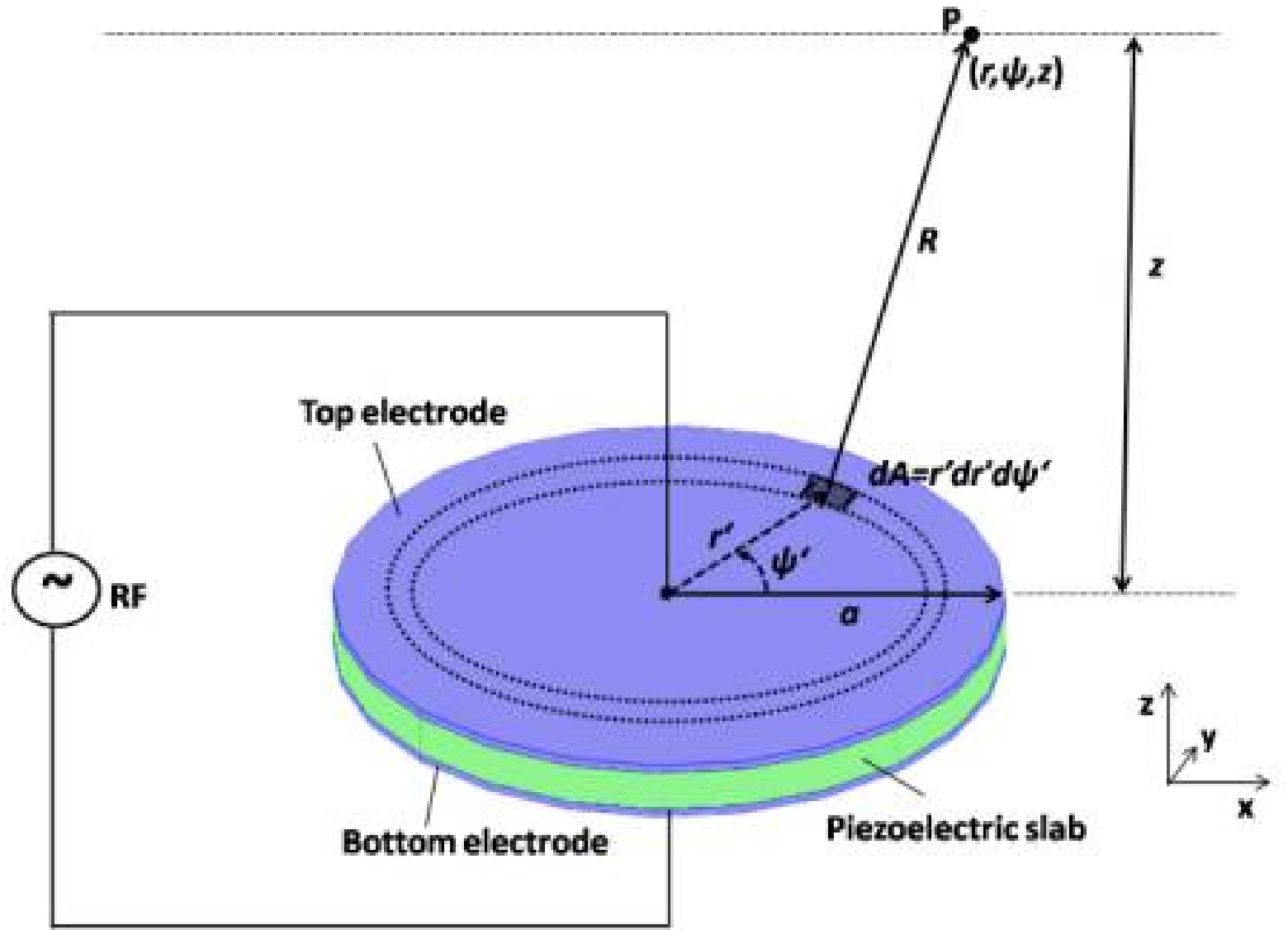} &
\includegraphics[width=0.9\columnwidth]{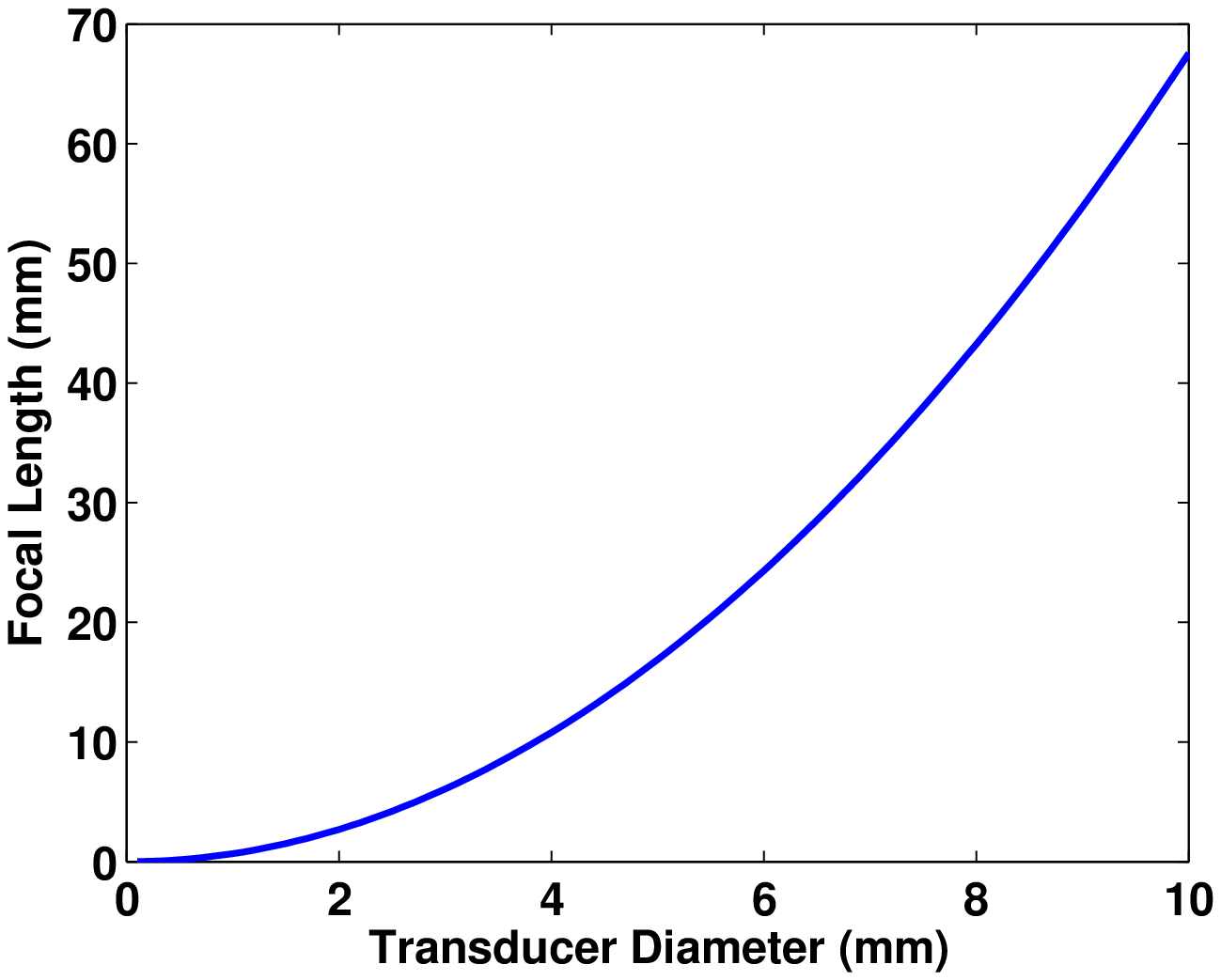}\\
(a) & (b)~\\
\end{tabular}
\end{center}
\vspace{-0.1in}
\caption{(a) Piston transducer with circular electrode shape (b) Natural focal length versus diameter of the transducer at 4MHz in water} 
\label{fig:CircXudcer}\vspace{-0.22in}
\end{figure*}

The rest of the paper is organized as follows. In section~\ref{sec:CirXducer}, we present the simulation model to compute acoustic potential and particle displacement due to circularly-symmetric piezoelectric transducers. In section~\ref{sec:ConcenRingXducer}, we present the design and analysis of circular concentric-ring transducer; we also discuss the limitations of such transducer. In section~\ref{sec:SectoredXducer}, we present the design of sectored annular transducers and compare the acoustic field due to transducers of different sector-angles. In this section, we also discuss the advantages of using $90^{\circ}$ transducers over other sector angles. Finally, we conclude the paper in section ~\ref{sec:conclusion}.

\section{Acoustic Field Computation Due to Circularly Symmetric Transducers}
\label{sec:CirXducer}
For a circular transducer, as shown in figure-\ref{fig:CircXudcer}, we calculate the displacement potential at any point P using full Rayleigh-Sommerfeld integral~\cite{GSKinoBook1987, RosenbaumBAWbook1988}. We assume that the particle displacement just above the surface of the transducer is $u_0=1$ for our simulations. Typically, $u_0$ varies with time because piezoelectric transducer is excited by a time varying RF-signal. To compute the acoustic potential due to whole circular electrode-area of radius $a$, we first consider a small elemental area, at a radial distance $r'$ and angular distance $\psi'$, on top circular-electrode. We then use integration to find the acoustic potential due to complete electrode area. Let the area of a small element is $dA = r'dr'd\psi'$; this differential area element is located at coordinate ($r',\psi', 0$) in cylindrical system. The $z=0$ plane represents the top electrode on the slab. The acoustic potential ($\Phi$) due to whole circular electrode at point P located at ($r,\psi, z$) is given by:

\begin{figure*}[t]
\begin{center}
\begin{tabular}{ccc}
\hspace{-0.1in}
\includegraphics[width=0.33\textwidth]{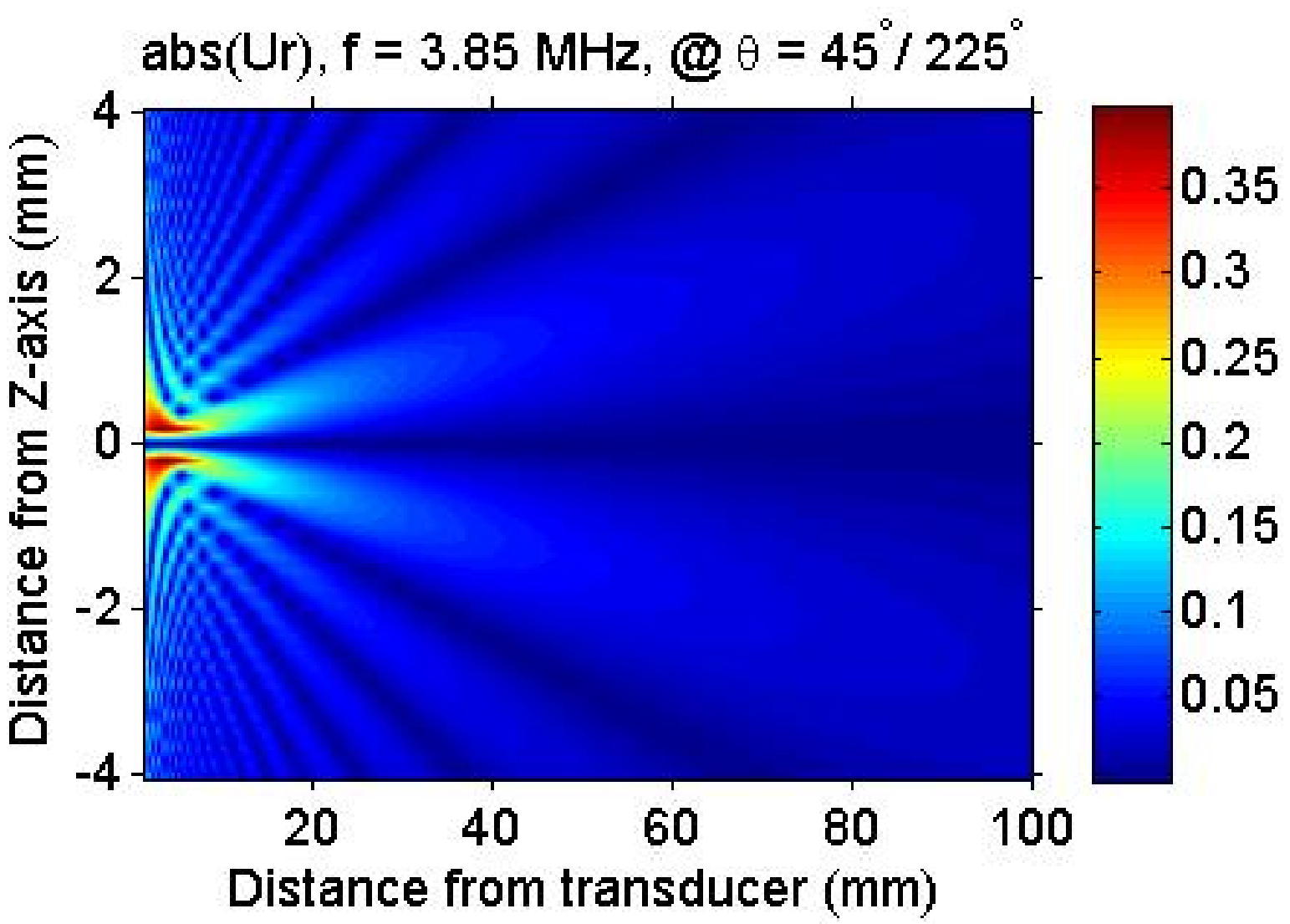}&
\hspace{-0.23in}
\includegraphics[width=0.33\textwidth]{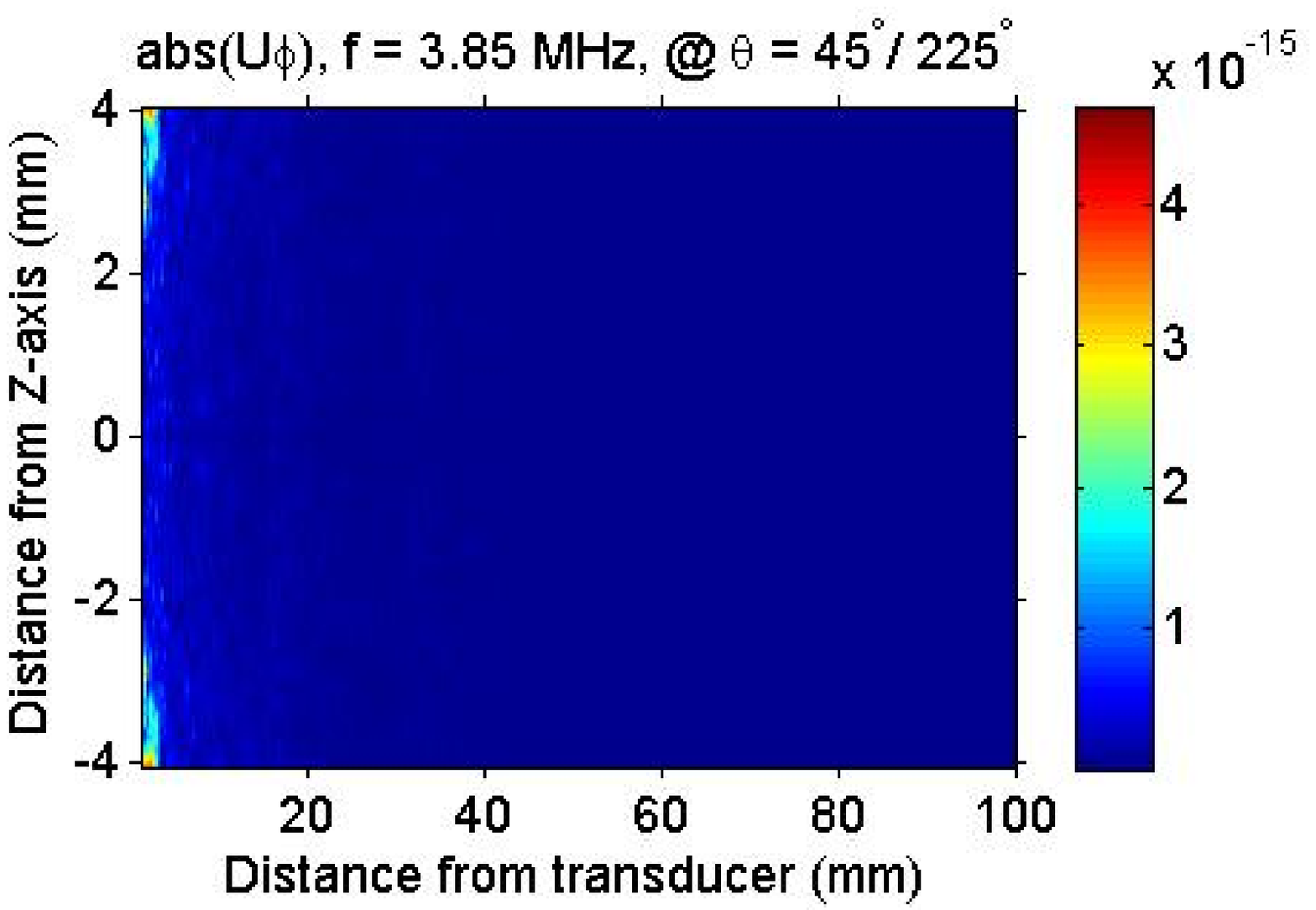}&
\hspace{-0.23in}
\includegraphics[width=0.33\textwidth]{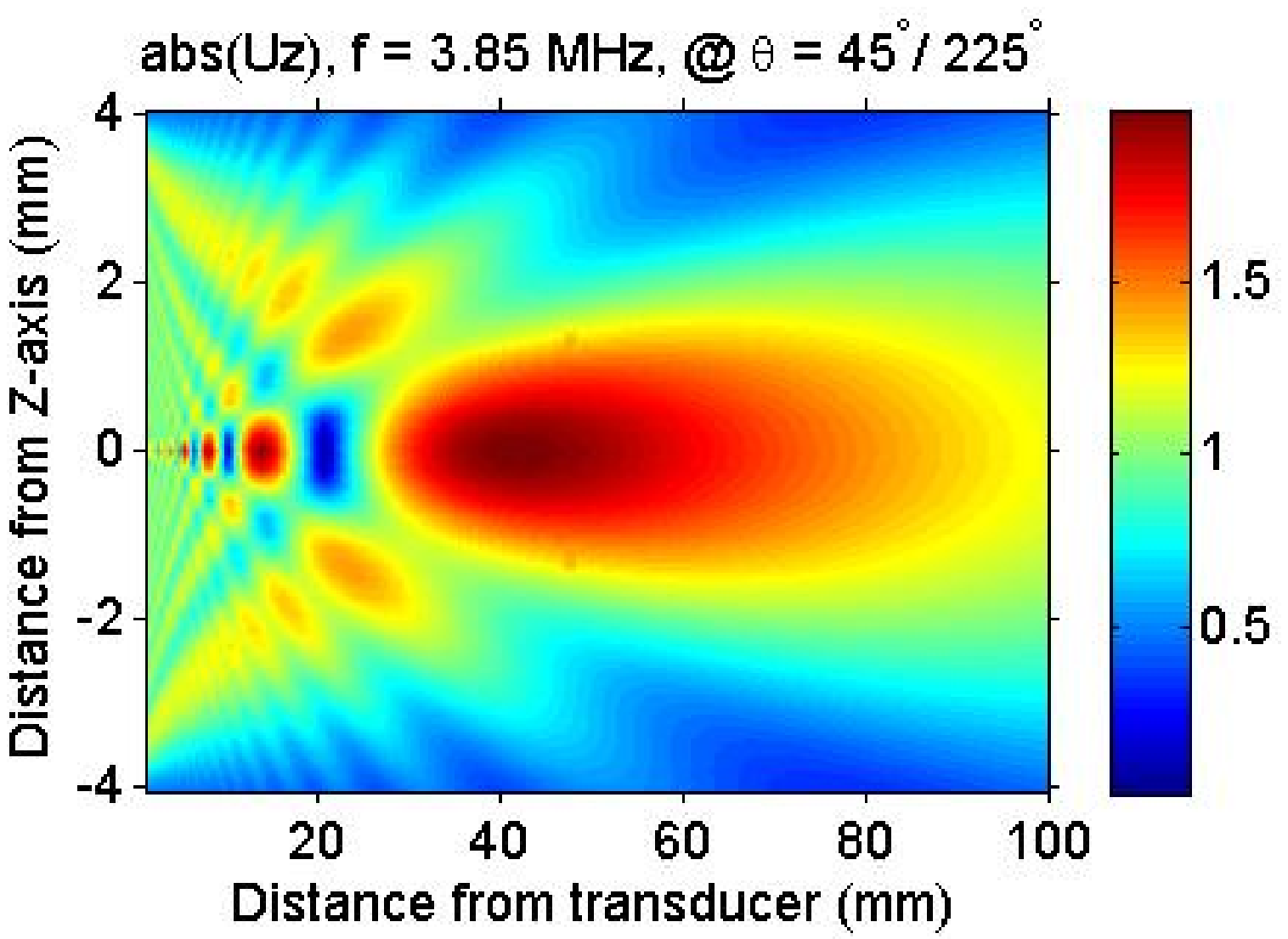}
\\
(a) & (b)& (c)\\
\end{tabular}
\end{center}
\vspace{-0.1in}
\caption{Acoustic particle displacement due to circular transducer of radius 4mm (a) radial component ($u_r$); (b) rotational component ($u_\psi$); (c) vertical component ($u_z$).} 
\label{fig:PartDispCircDisc}
\vspace{-0.22in}
\end{figure*}

\begin{equation}
\label{eq:PhiBasic}
\Phi\left(r,\psi,z\right) = -\frac{u_0}{2\pi} \int_{\psi'=0}^{2\pi} \int_{r'=0}^{a} \frac{e^{-\left(\alpha +jk \right)R}} {R} r'd\psi' dr'
\end{equation}

Where, $\alpha$ is the acoustic attenuation constant of the medium, $k$ is the wave number ($= 2\pi/\lambda$) and $R$ is the distance between point P and the elemental area $dA$. It can be shown that $R=\sqrt{z^2+r^2+r'^2-2rr'cos\left(\psi - \psi' \right)}$. 

Once the acoustic potential is known at point P, the relative particle displacement (in radial, vertical, and circumferential directions) can be calculated by differentiating the acoustic potential at that point. That is:

\begin{equation}
\label{eq:u_vs_Phi}
u = \nabla \Phi \left( r, \psi, z \right) = \left( \frac{\partial} {\partial r} \hat{r} + \frac{\partial} {r\partial \psi} \hat{\psi} + \frac{\partial} {\partial z} \hat{z} \right) \Phi \left( r, \psi, z \right)
\end{equation}

Hence, the different components of relative particle displacement, radial ($u_r$), circumferential ($u_\psi$), and vertical ($u_z$) can be derived as following:

\begin{align}
\label{eq:Ur}
\begin{split}
u_r = \frac{u_0}{2\pi} \int_{\psi'=0}^{2\pi} \int_{r'=0}^{a} \frac{e^{-\left(\alpha +jk \right)R}} {R^3} \big[1+\left(\alpha +jk \right)R \big].\\
 \big[ r-r'cos \left(\psi - \psi' \right) \big] r'd\psi' dr'
\end{split}
\end{align}

\begin{align}
\label{eq:Upsi}
\begin{split}
u_\psi = \frac{u_0}{2\pi} \int_{\psi'=0}^{2\pi} \int_{r'=0}^{a} \frac {e^{-\left(\alpha +jk \right)R}} {R^3} \big[1+\left(\alpha +jk \right)R \big].\\
 r'^2 sin \left(\psi - \psi' \right) d\psi' dr'
\end{split}
\end{align}

\begin{align}
\label{eq:Uz}
\begin{split}
u_z = \frac{u_0}{2\pi} \int_{\psi'=0}^{2\pi} \int_{r'=0}^{a} \frac{e^{-\left(\alpha +jk \right)R}} {R^3} \big[1+\left(\alpha +jk \right)R \big].\\
 zr'd\psi' dr'
\end{split}
\end{align}

The radial, circumferential and vertical components of particle displacement for a circular transducer (radius = 4mm) at different z-heights are given in figure~\ref{fig:PartDispCircDisc}. It could be observed from the figure that the circumferential component of particle displacement is almost zero for a circular transducer, mainly because of its axial symmetry; only z-component is prominent and the maxima/focus is at 41.2mm when the transducer is excited at 3.85MHz and is placed in water as medium. Such a transducer is widely used for characterizing the crystal structure in solids and for medical ultrasounds, but is not good for microfluid processing applications.

\begin{figure}[b]
\begin{center}
\begin{tabular}{cc}
\includegraphics[width=0.25\columnwidth]{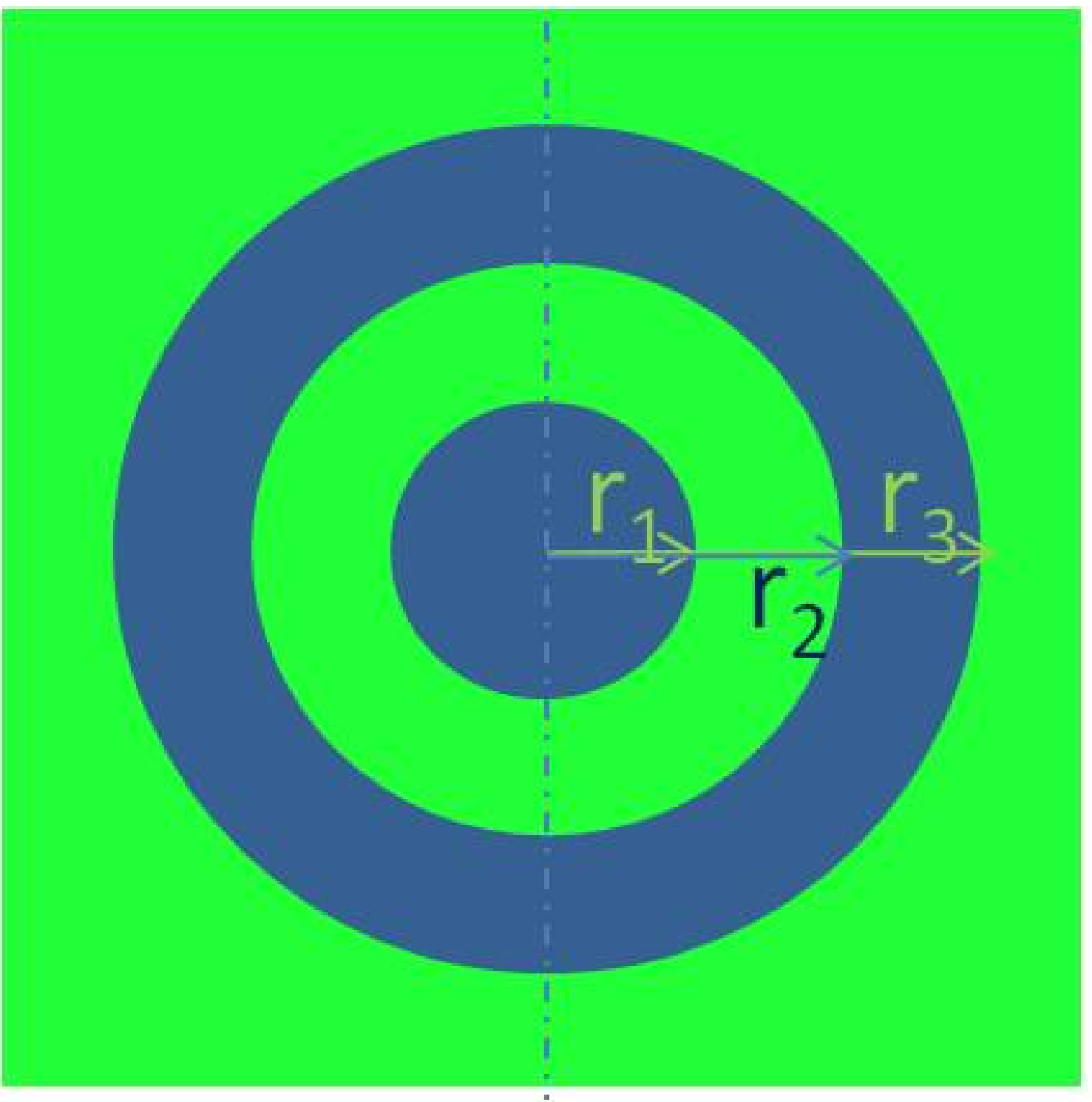} &
\includegraphics[width=0.46\columnwidth]{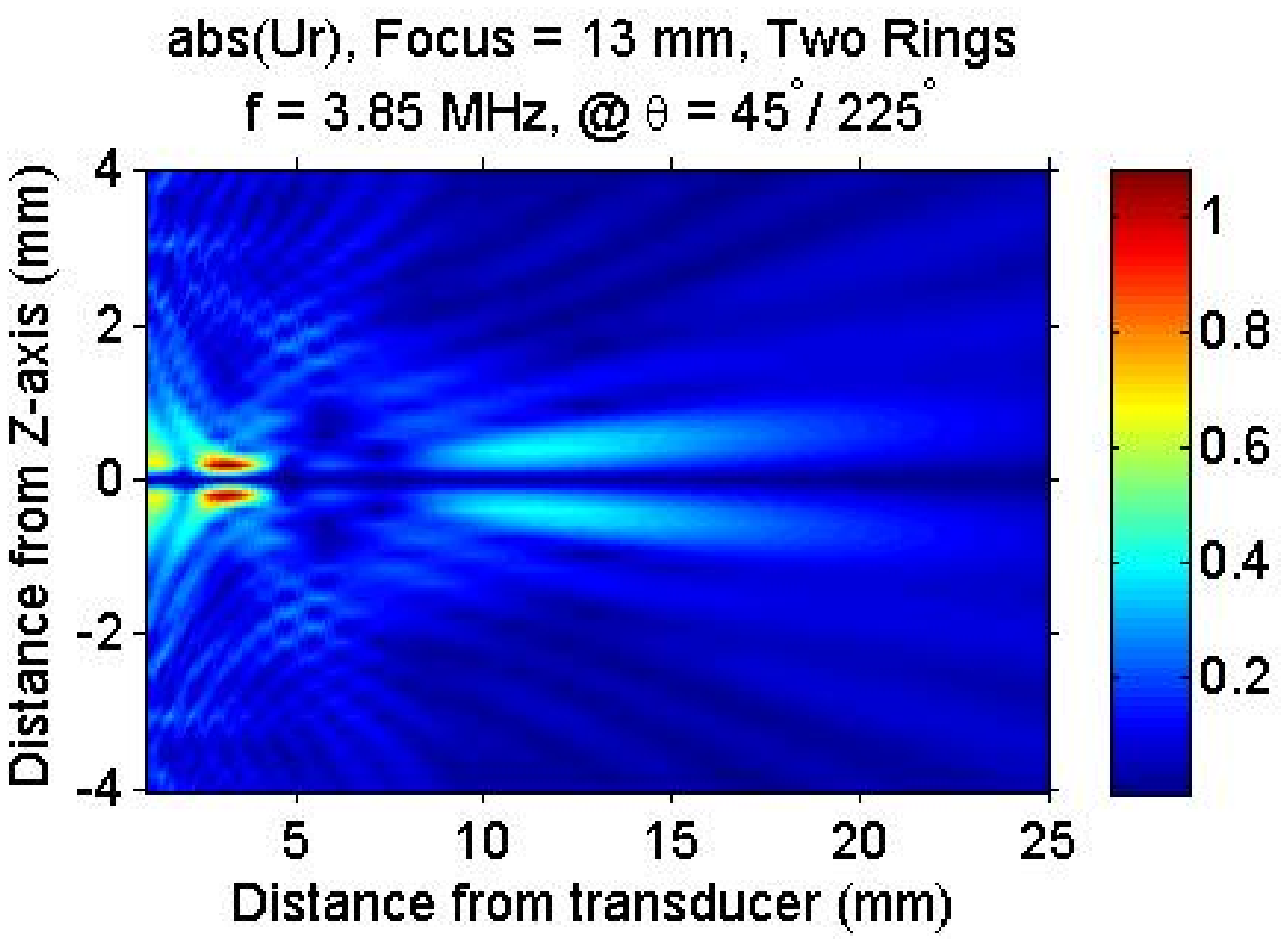}\\
(a) & (b)\\
\includegraphics[width=0.46\columnwidth]{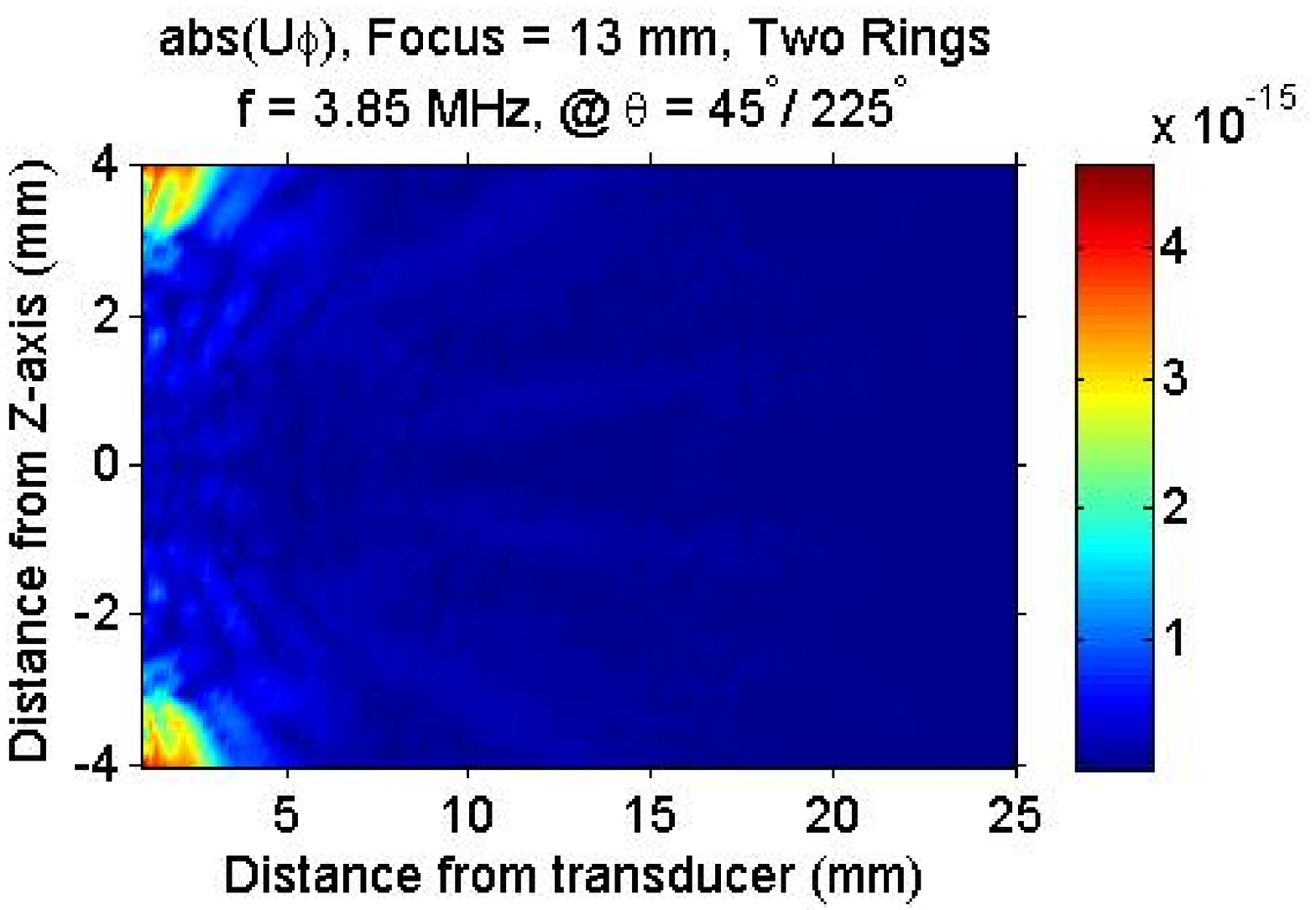} &
\includegraphics[width=0.46\columnwidth]{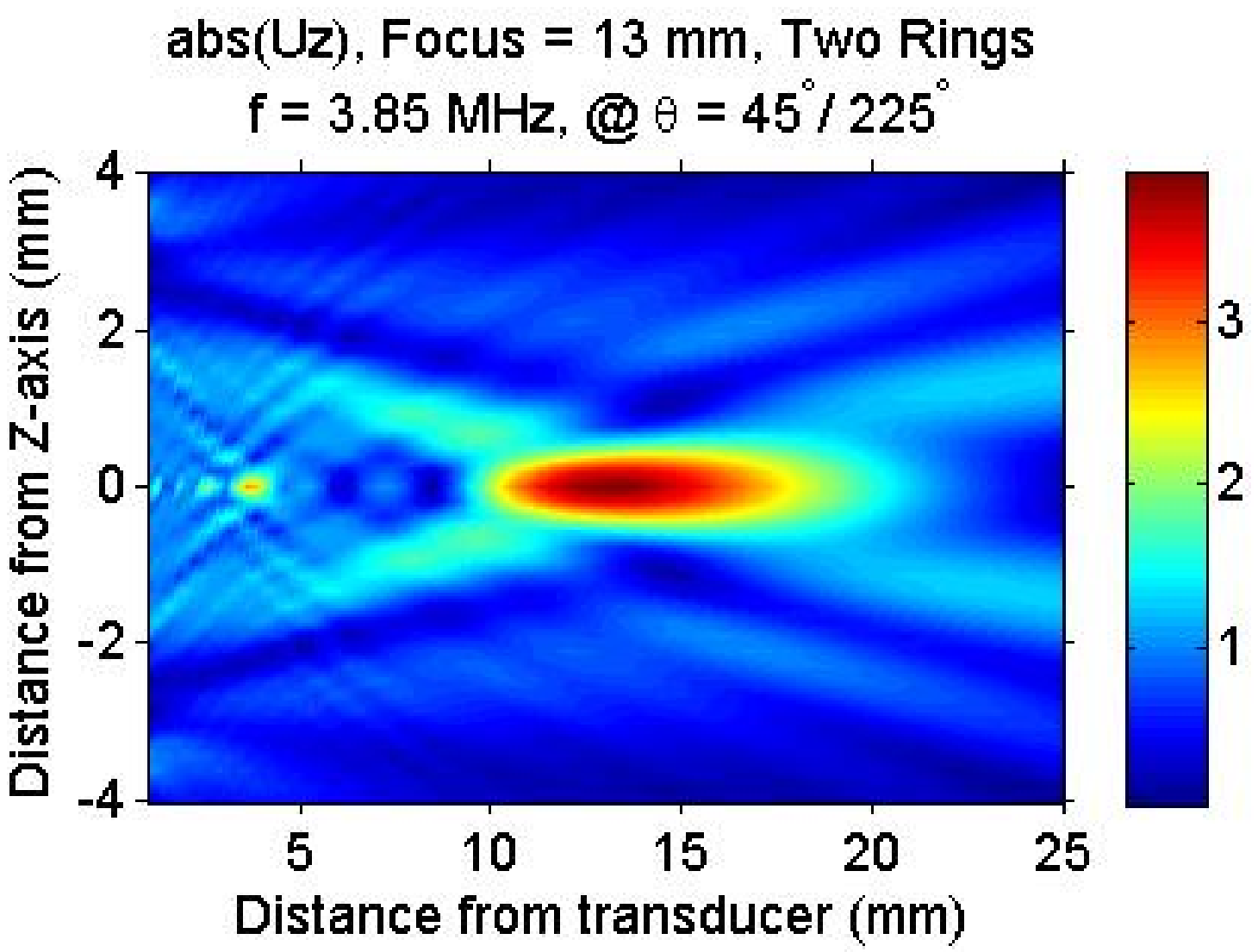}\\
(c) & (d)\\
\end{tabular}
\end{center}
\vspace{-0.1in}
\caption{(a) Circular ring-shaped transducer (2-rings); Acoustic particle displacement intensity for the transducer designed at FL=13mm: (b) radial component ($u_r$), (c) circumferential component ($u_\psi$), (d) vertical component ($u_z$).} 
\label{fig:PartDisp360Deg2Rings}
\vspace{-0.3in}
\end{figure}

\section{Piezoelectric Transducer with Circular Concentric Ring Shaped Electrodes}
\label{sec:ConcenRingXducer}
To better control the focal length of an ultrasonic transducer, while allowing the transducer to be of any size, the electrodes on the transducer could be 
designed in the shape of rings. Such an ultrasonic transducer is equivalent to a Fresnel focusing lens in optics. Depending on the design of inner most ring, there are two types of Fresnel lens, positive source and negative source. When the inner radius of first ring is zero, the transducer is termed as negative source Fresnel lens and when the inner radius of the first ring is non-zero, the transducer is called as positive source. In other words, the electrode-area for a positive-source transducer is the non-electrode area for a negative-source transducer and vice-versa. The ring-radii for such a transducer is calculated using the below formula~\cite{Xzhu98, HuangMEMS2001}:
\vspace{-0.2in}

\begin{equation}
\label{eq:RingRadii}
r_n = \sqrt{\left( n\frac{\lambda}{2} \right)^2 + n\lambda F}
\end{equation}

Where, n is an integer (0,1,...) representing inner/outer radii rings, $\lambda$ is the wavelength of ultrasound wave in medium and F is the focal length. For ring-shaped transducer also, the particle displacements at different point in space are computed using equations~(\ref{eq:Ur})-(\ref{eq:Uz}); however, the integral limits of $r'$ are taken such that all the rings are covered during integration. 

By comparing the plots of figure-\ref{fig:PartDispCircDisc} and \ref{fig:PartDisp360Deg2Rings}, we confirm that the ring-shaped electrode has higher radial and vertical component of particle displacement than that due to the circular transducer even though the total active area is higher in a circular transducer. The maximum radial particle displacement is 0.39 (a.u.) in circular transducer case, while it is 1.1 (a.u.) in the case of 360-degree ring shaped transducer built in the same footprint-area. Similarly, the maximum vertical component of particle displacement due to circular and ring-shaped transducers are 1.9 (a.u.) and 3.9 (a.u.) respectively. The circumferential component due to both circular and ring-shaped transducer is negligible. Next, we present some transducer designs which have higher rotational component of acoustic particle displacement as well as better radial and vertical component of particle displacement per unit electrode area.

\section{Piezoelectric Transducer with Sectored Annular Electrodes}
\label{sec:SectoredXducer}
\begin{figure}[b]
\begin{center}
\begin{tabular}{cccc}
\includegraphics[width=0.1\textwidth]{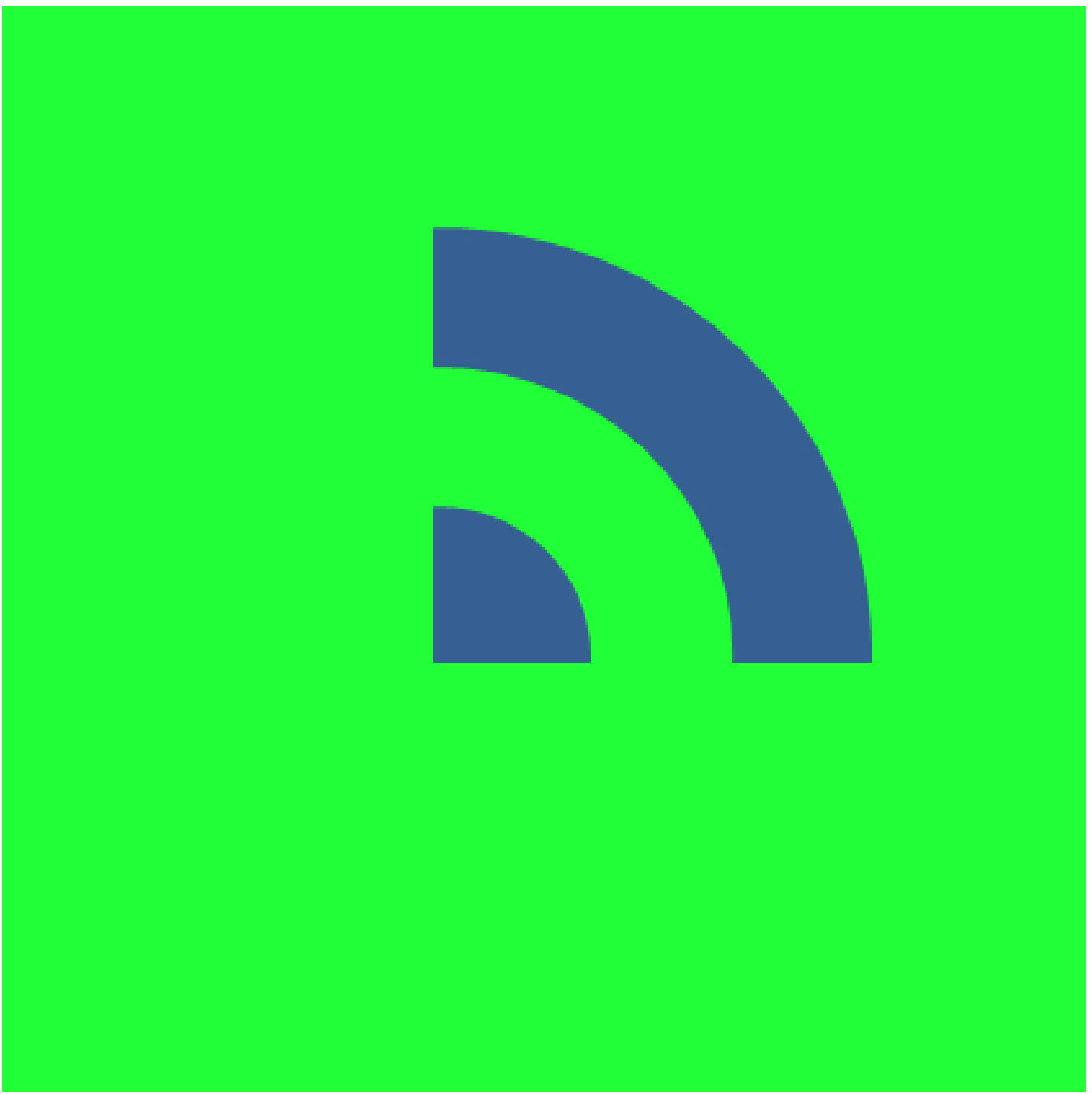} &
\includegraphics[width=0.1\textwidth]{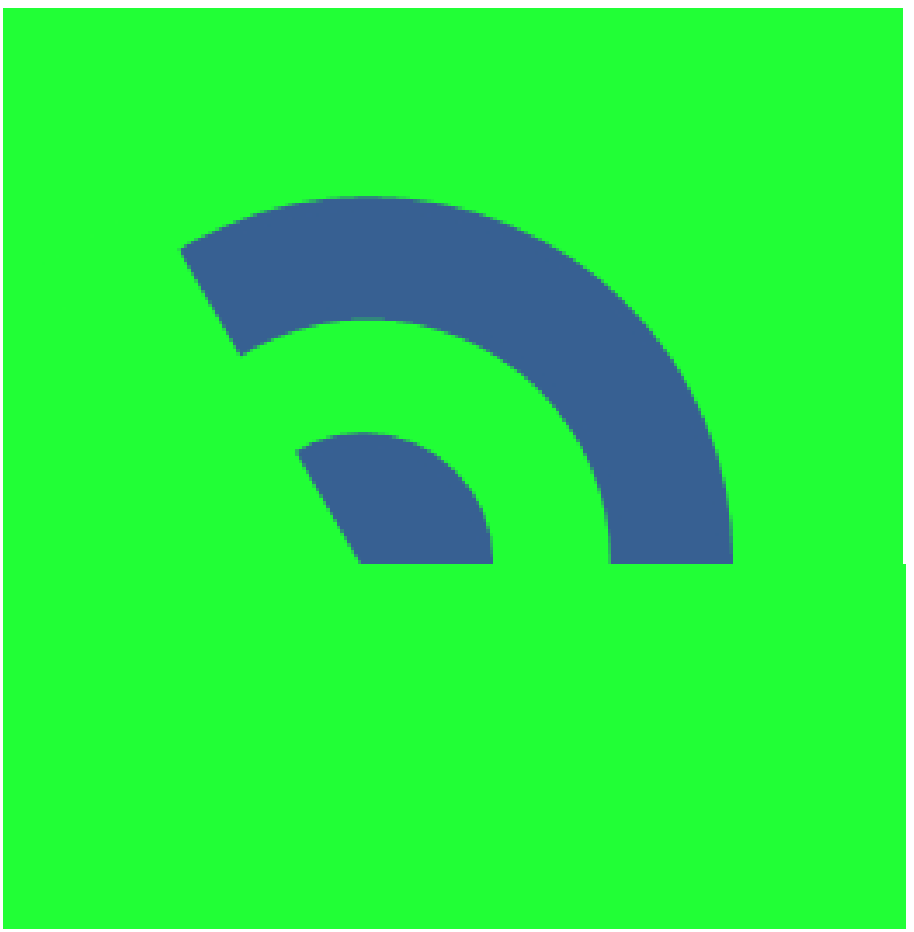} &
\includegraphics[width=0.1\textwidth]{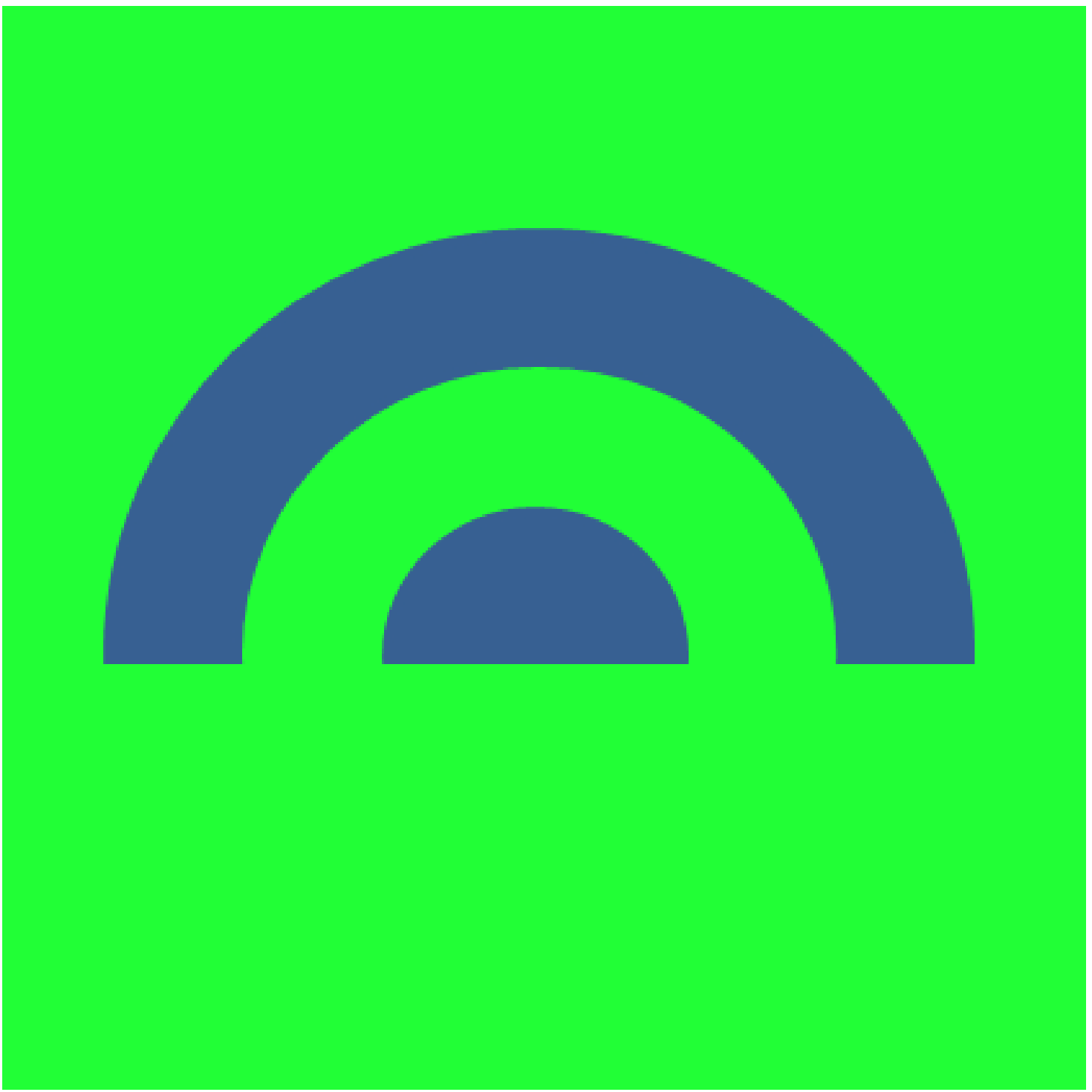} &
\includegraphics[width=0.1\textwidth]{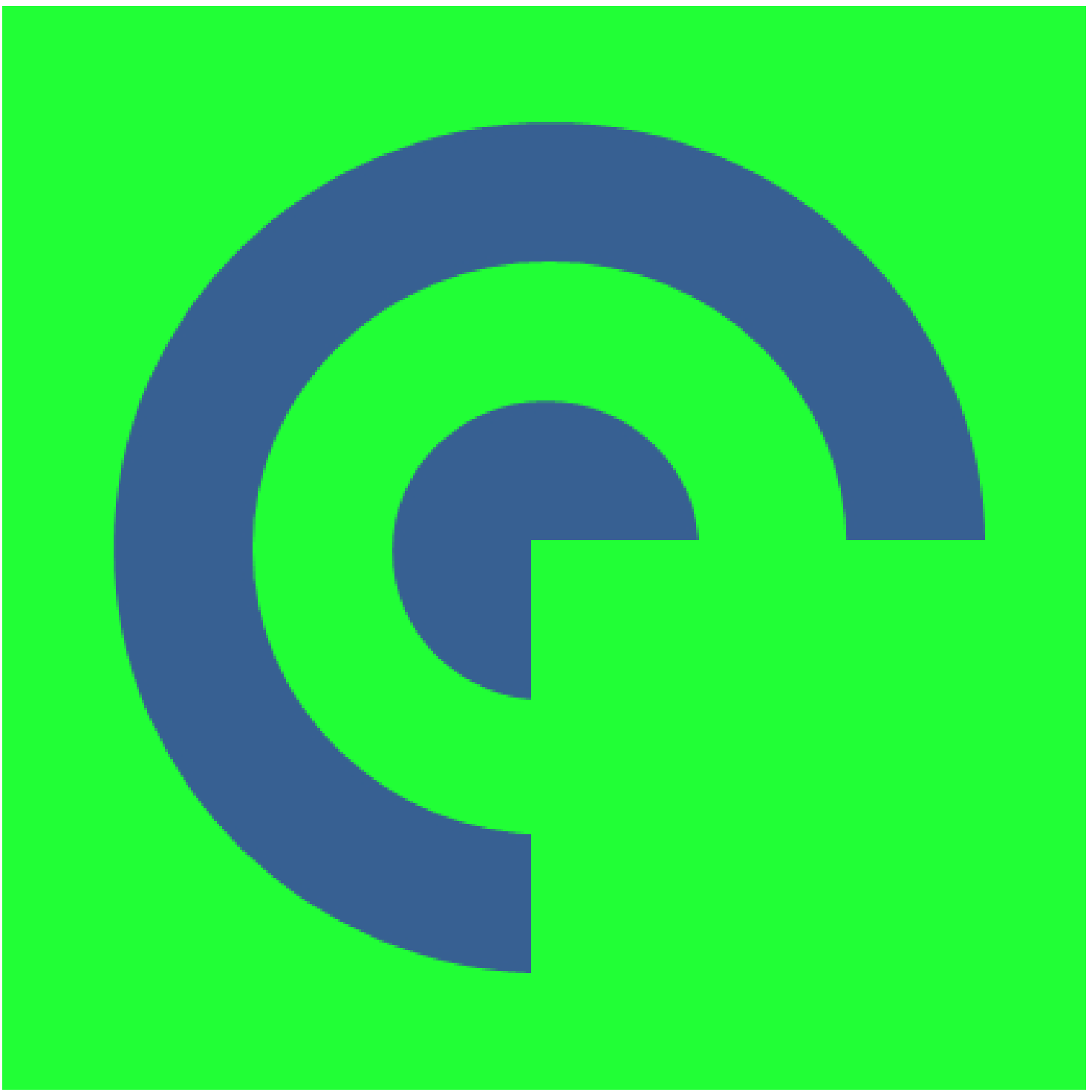}\\
(a) & (b) & (c) & (d)\\
\end{tabular}
\end{center}
\vspace{-0.1in}
\caption{Sectored Annular Transducers: (a) $90^{\circ}$; (b) $180^{\circ}$, (c) $270^{\circ}$.} 
\label{fig:SectoredXducers}
\vspace{-0.22in}
\end{figure}

Since both circular disc and $360^{\circ}$ ring transducers have circular symmetry, the rotational component of acoustic field is zero. We design non-$360^{\circ}$ annular transducers to increase the circumferential (responsible for vortexing) component of particle displacement in fluid due to acoustic potential. We present simulation results for four sector-angles other than $360^{\circ}$ angle: $90^{\circ}$, $120^{\circ}$, $180^{\circ}$, and $270^{\circ}$; these transducers are shown in figure-\ref{fig:SectoredXducers}. The $90^{\circ}$ transducer was first proposed in~\cite{VibhuMEMS2000} and it was shown that such a transducer generates rotational field causing the liquid to vortex in opposite directions.

\begin{figure*}[tb]
\begin{center}
\begin{tabular}{cccc}
\includegraphics[width=0.47\columnwidth]{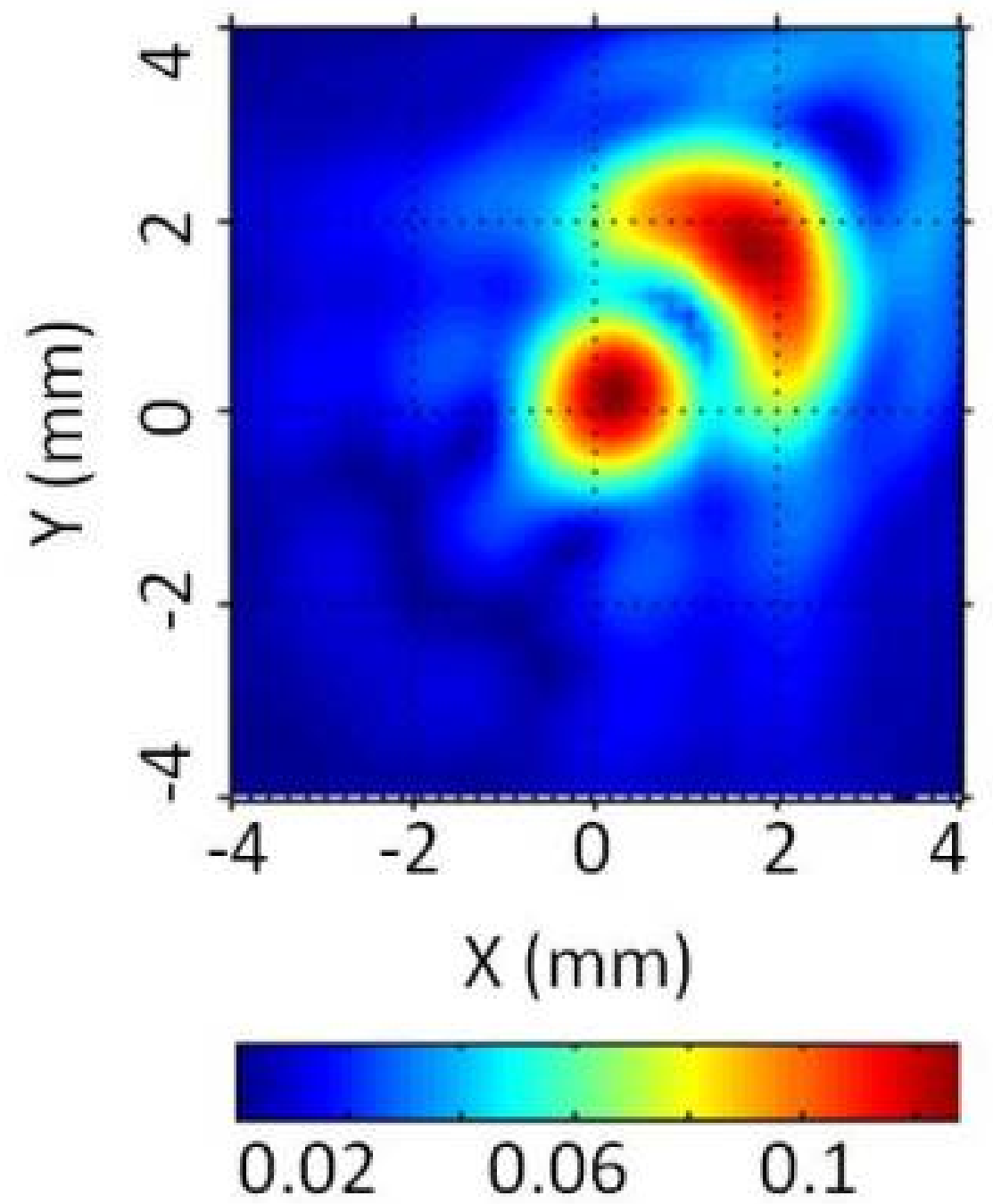} &
\hspace{-0.1in}
\includegraphics[width=0.47\columnwidth]{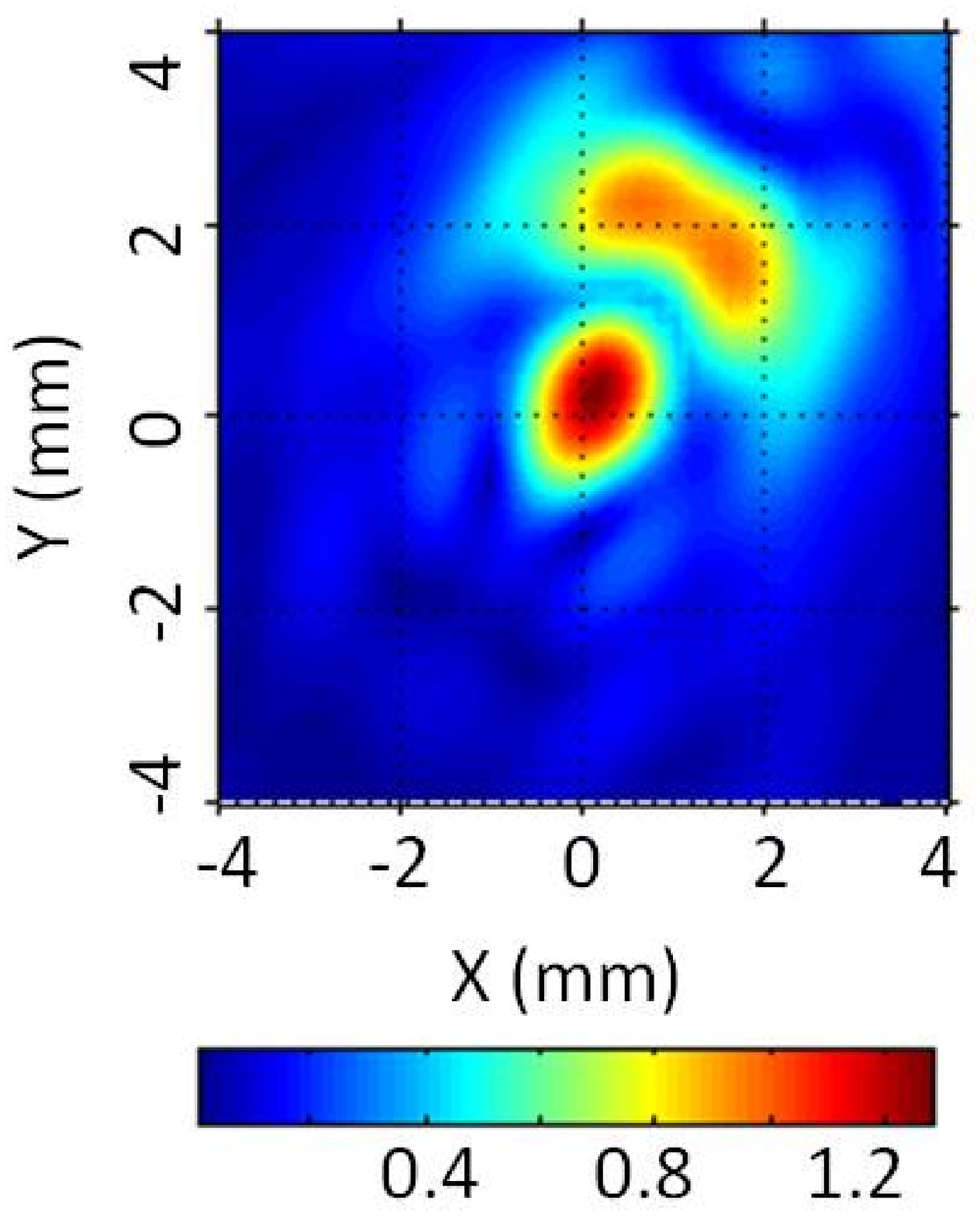} &
\includegraphics[width=0.47\columnwidth]{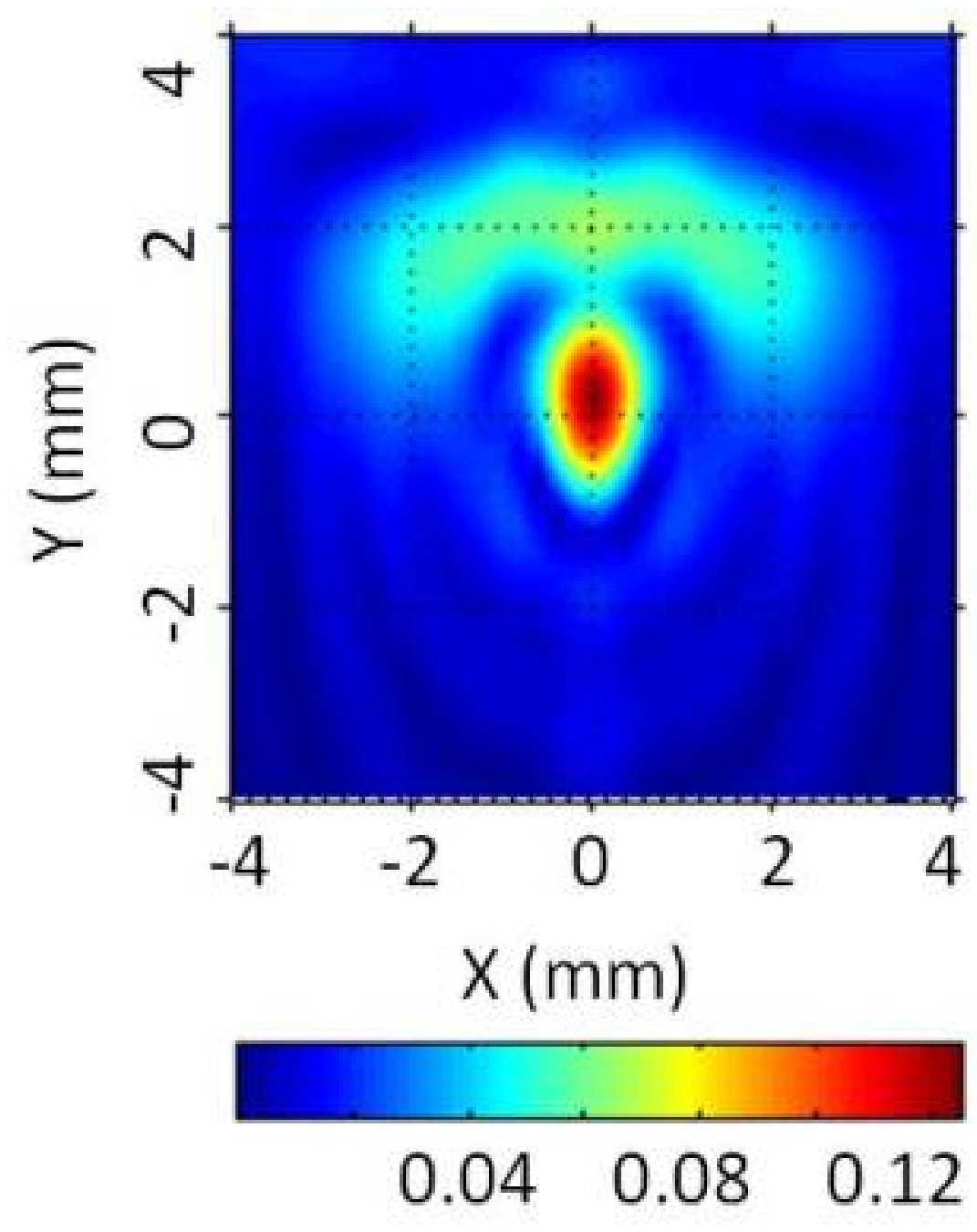} &
\hspace{-0.1in}
\includegraphics[width=0.47\columnwidth]{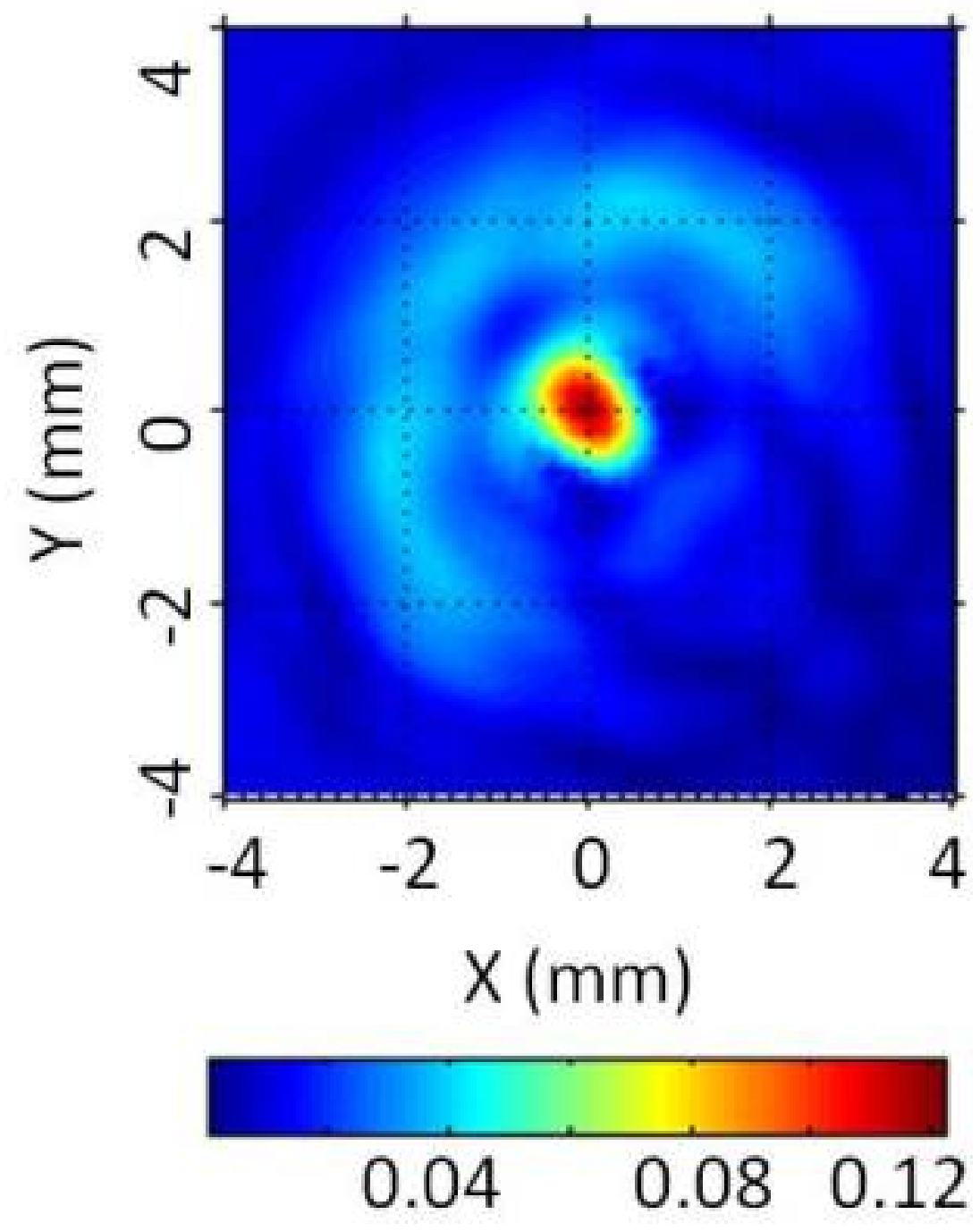}\\
(a) & (b) & (c) & (d)\\
\includegraphics[width=0.47\columnwidth]{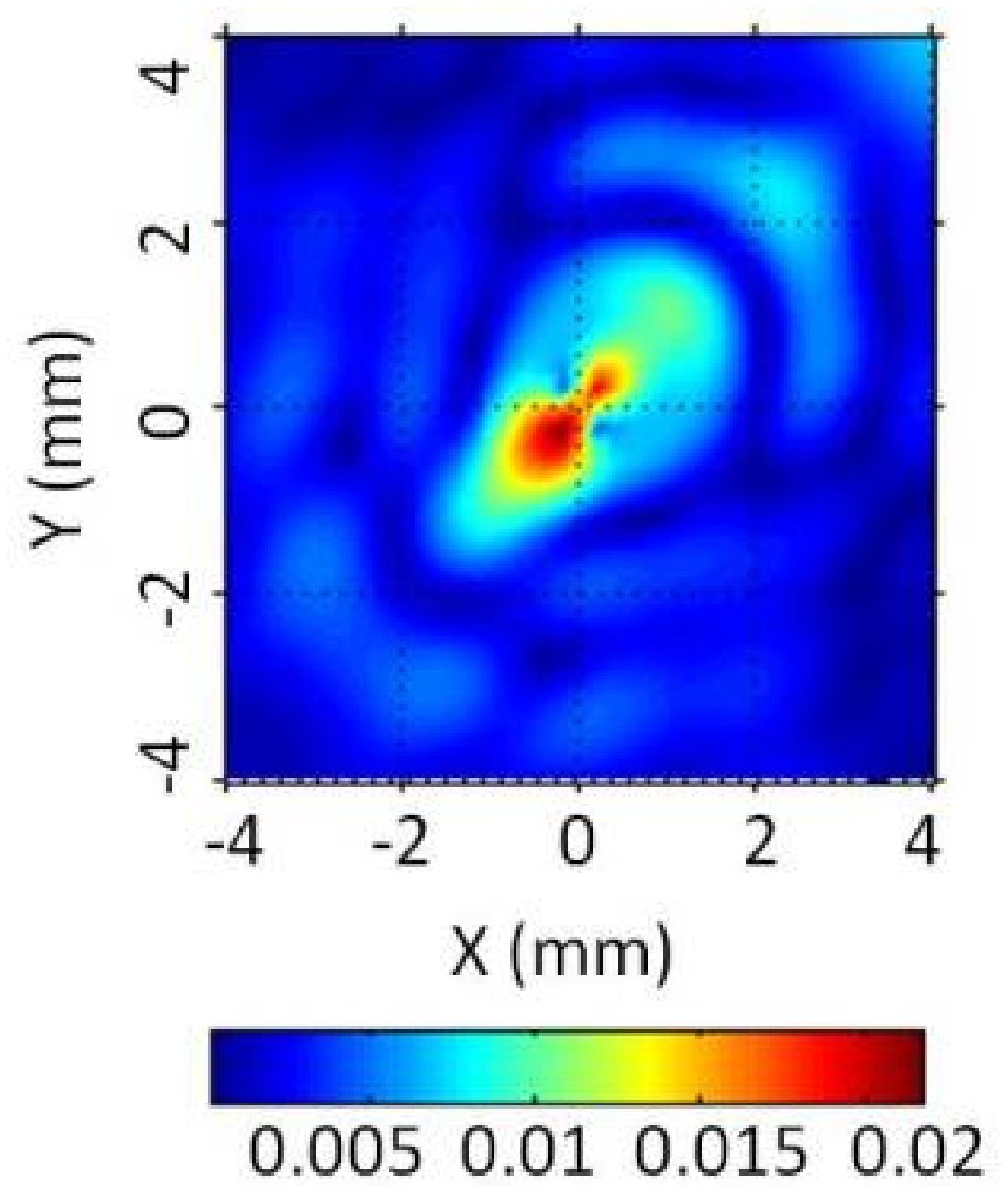} &
\hspace{-0.1in}
\includegraphics[width=0.47\columnwidth]{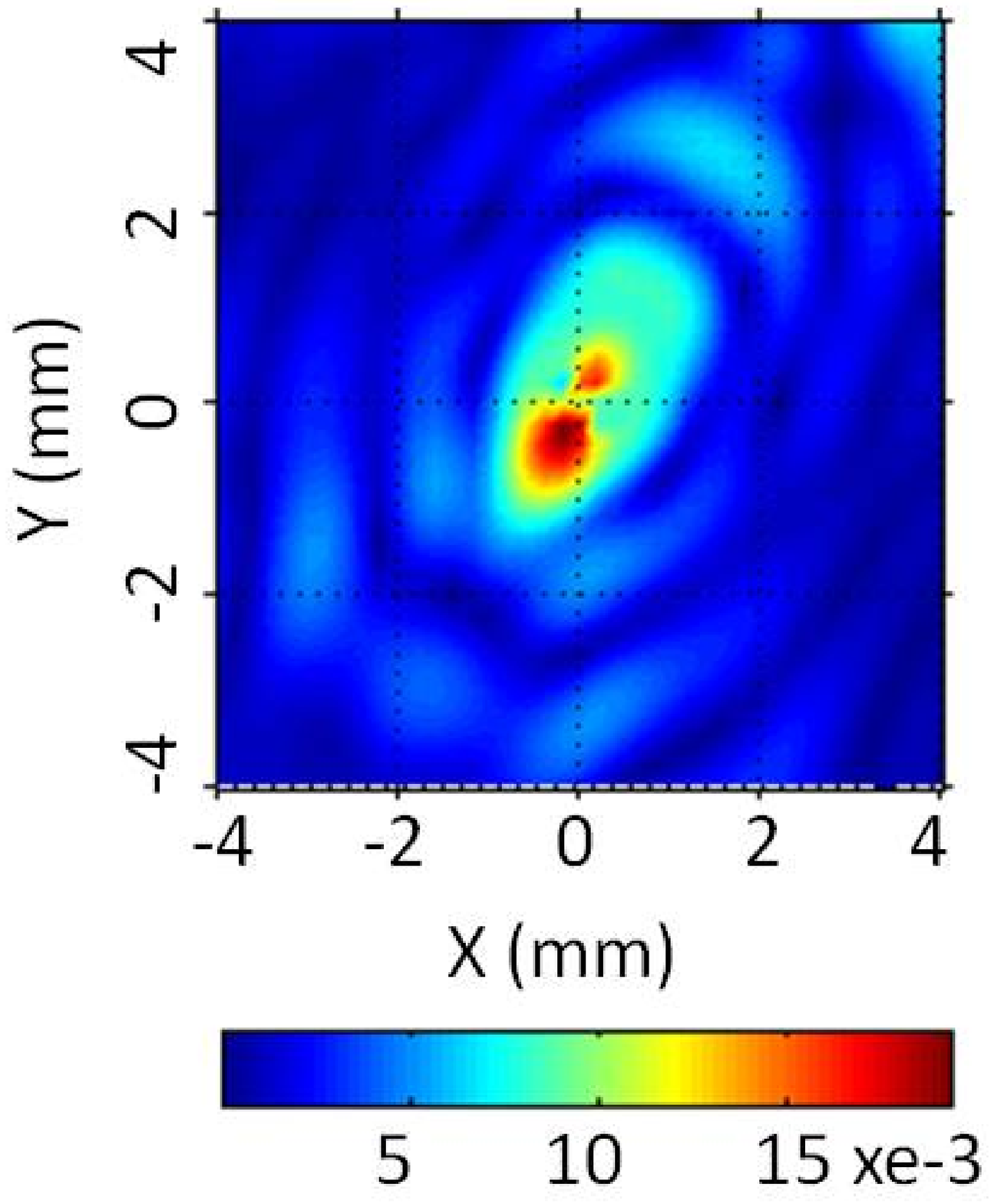} &
\includegraphics[width=0.47\columnwidth]{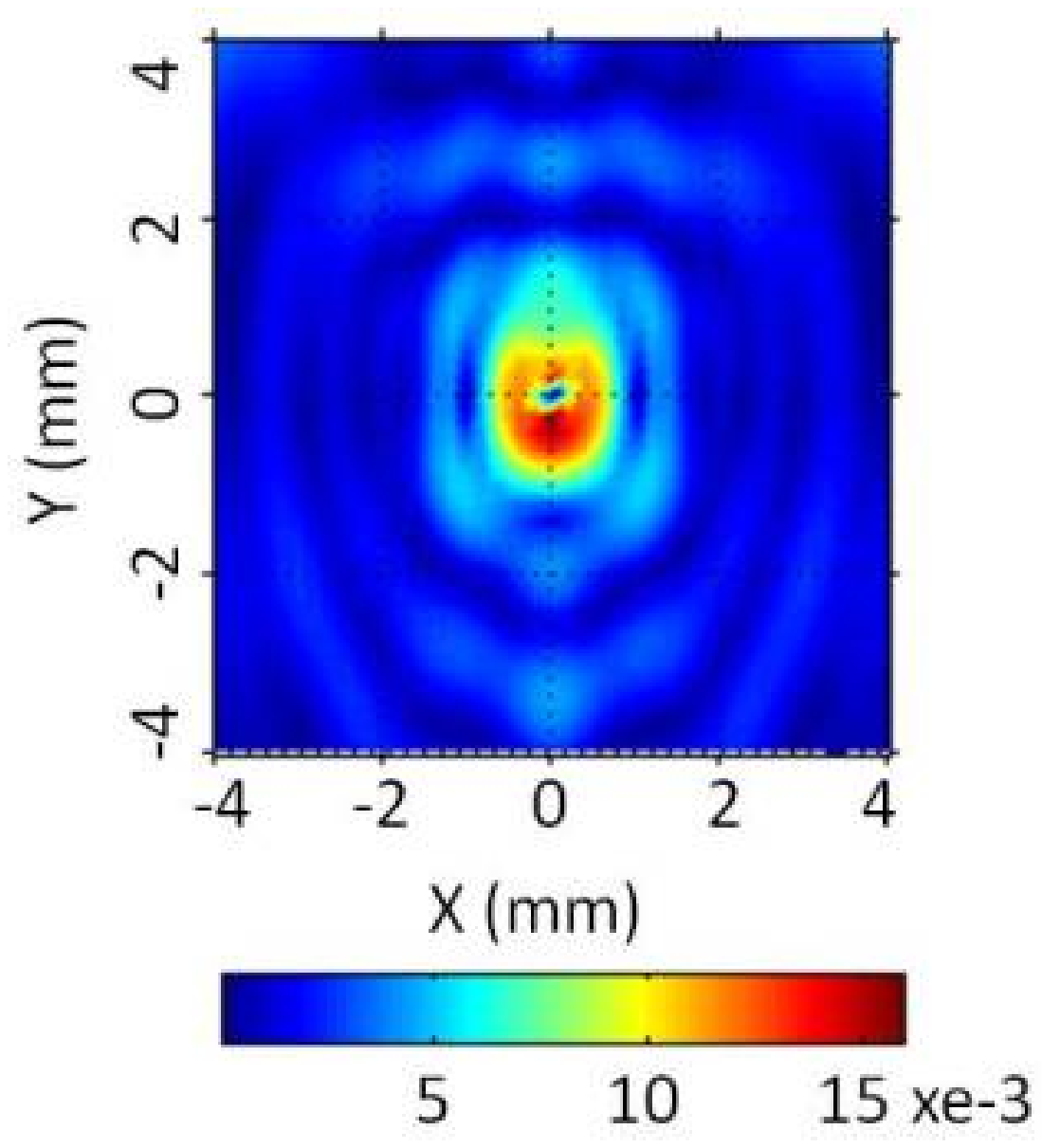} &
\hspace{-0.1in}
\includegraphics[width=0.47\columnwidth]{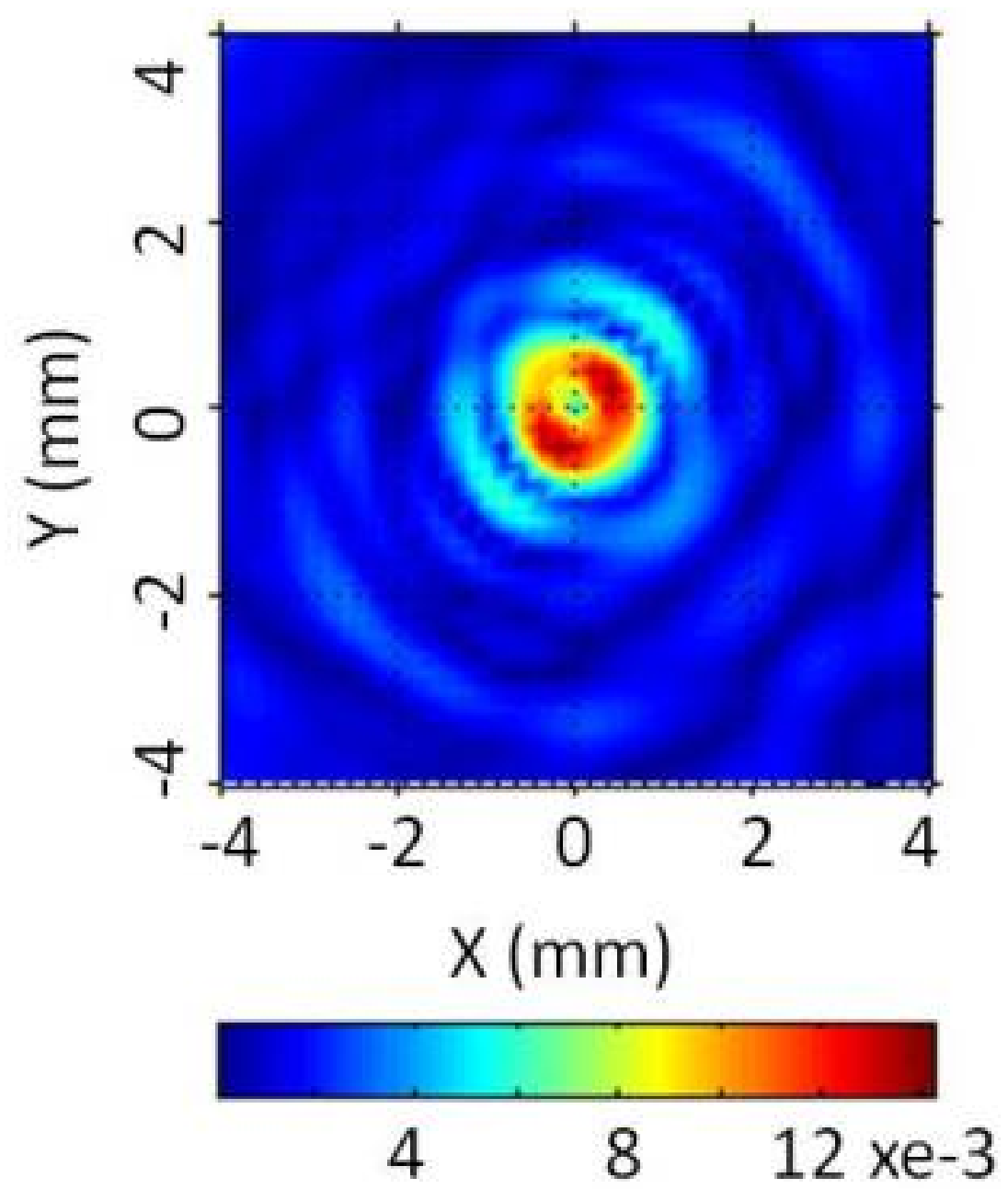}\\
(e) & (f) & (g) & (h)\\
\includegraphics[width=0.47\columnwidth]{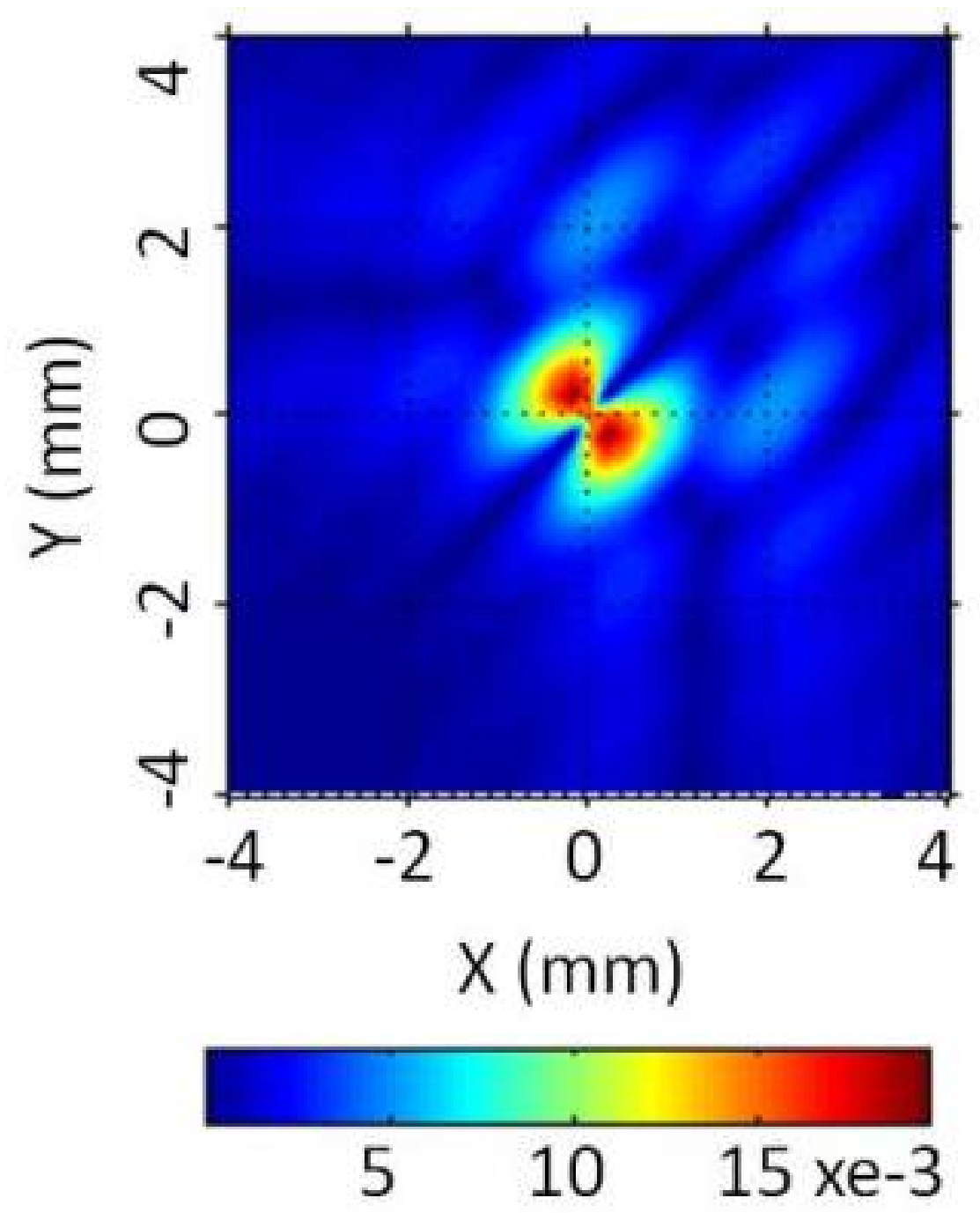} &
\hspace{-0.1in}
\includegraphics[width=0.47\columnwidth]{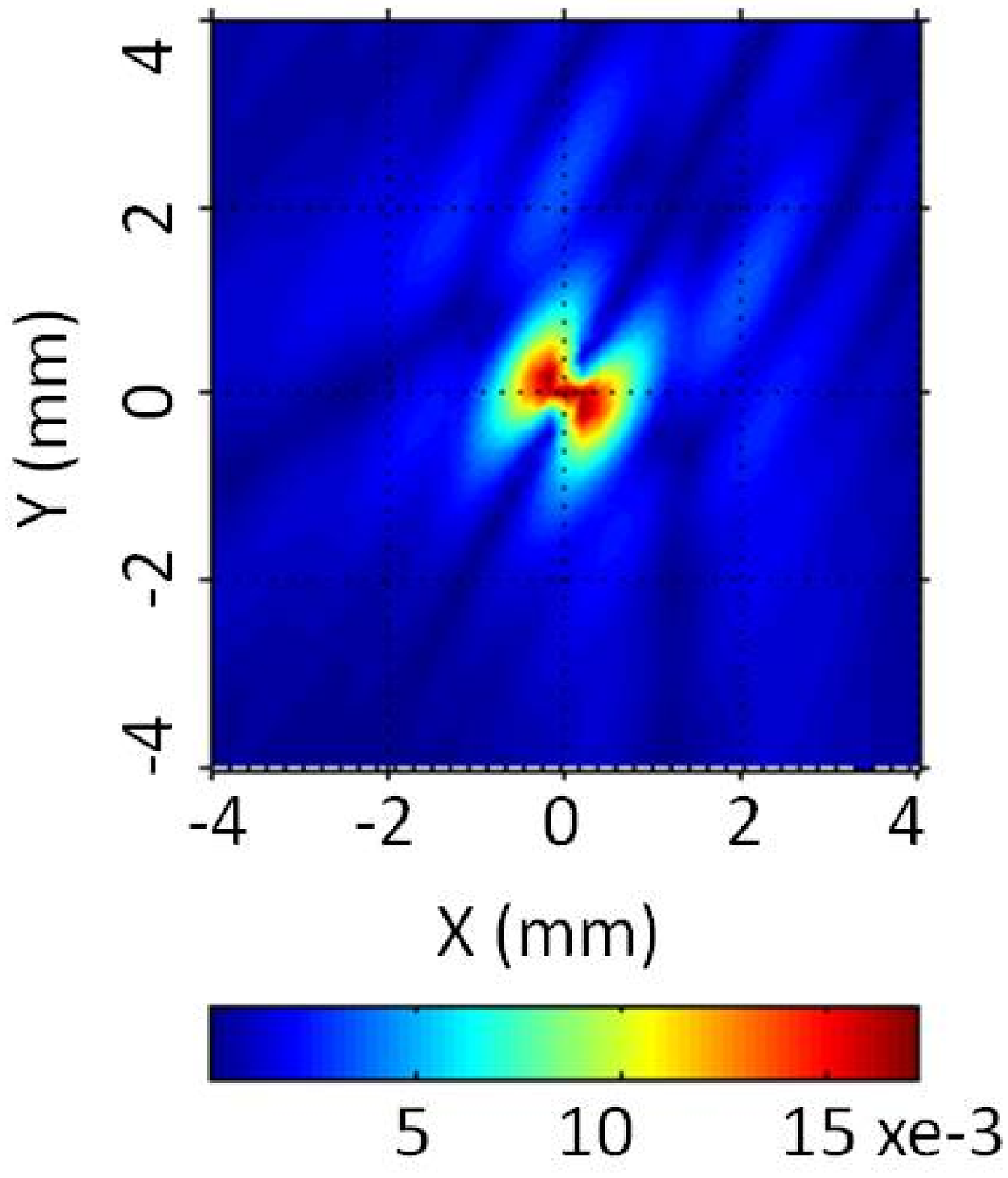} &
\includegraphics[width=0.47\columnwidth]{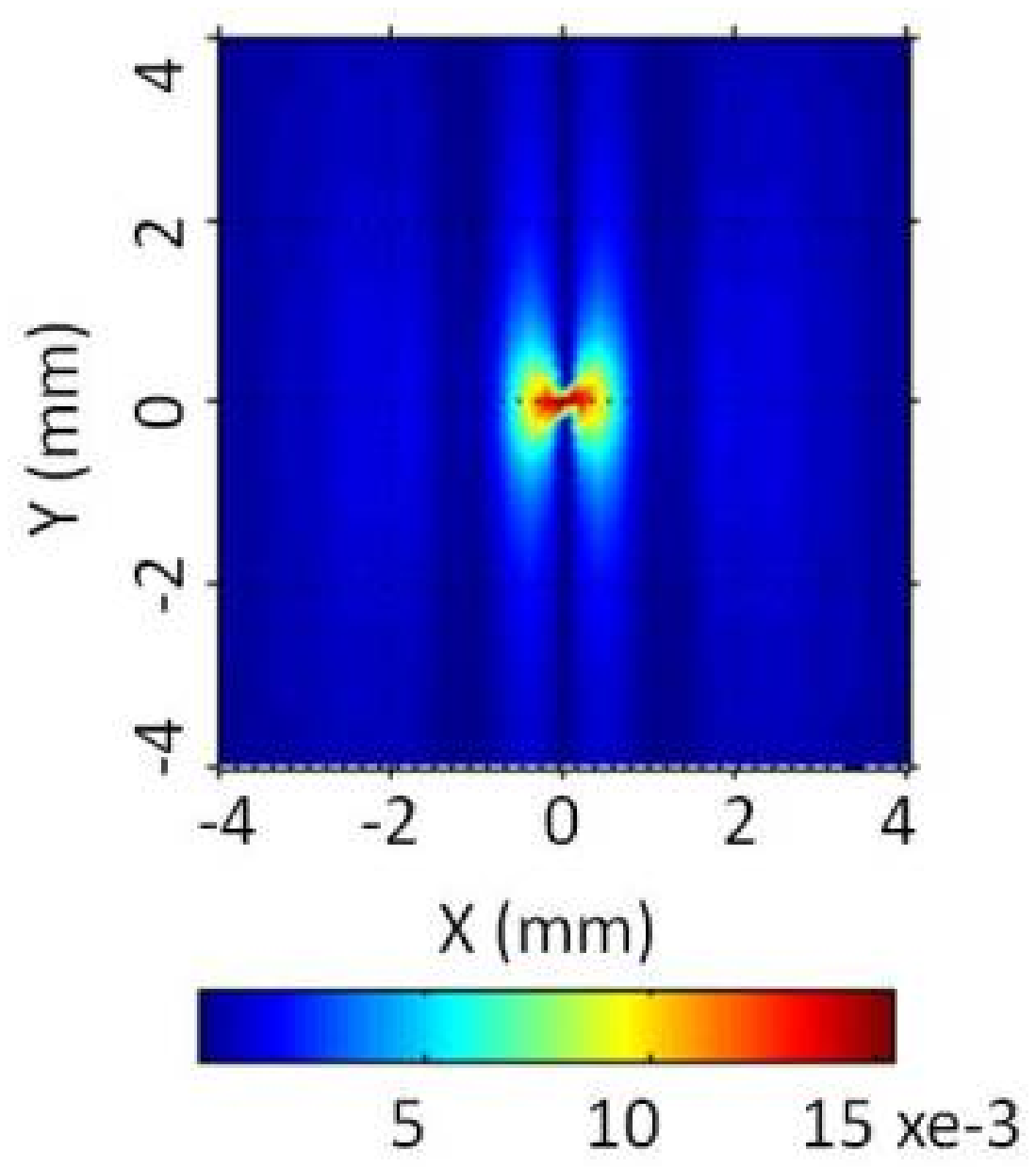} &
\hspace{-0.1in}
\includegraphics[width=0.47\columnwidth]{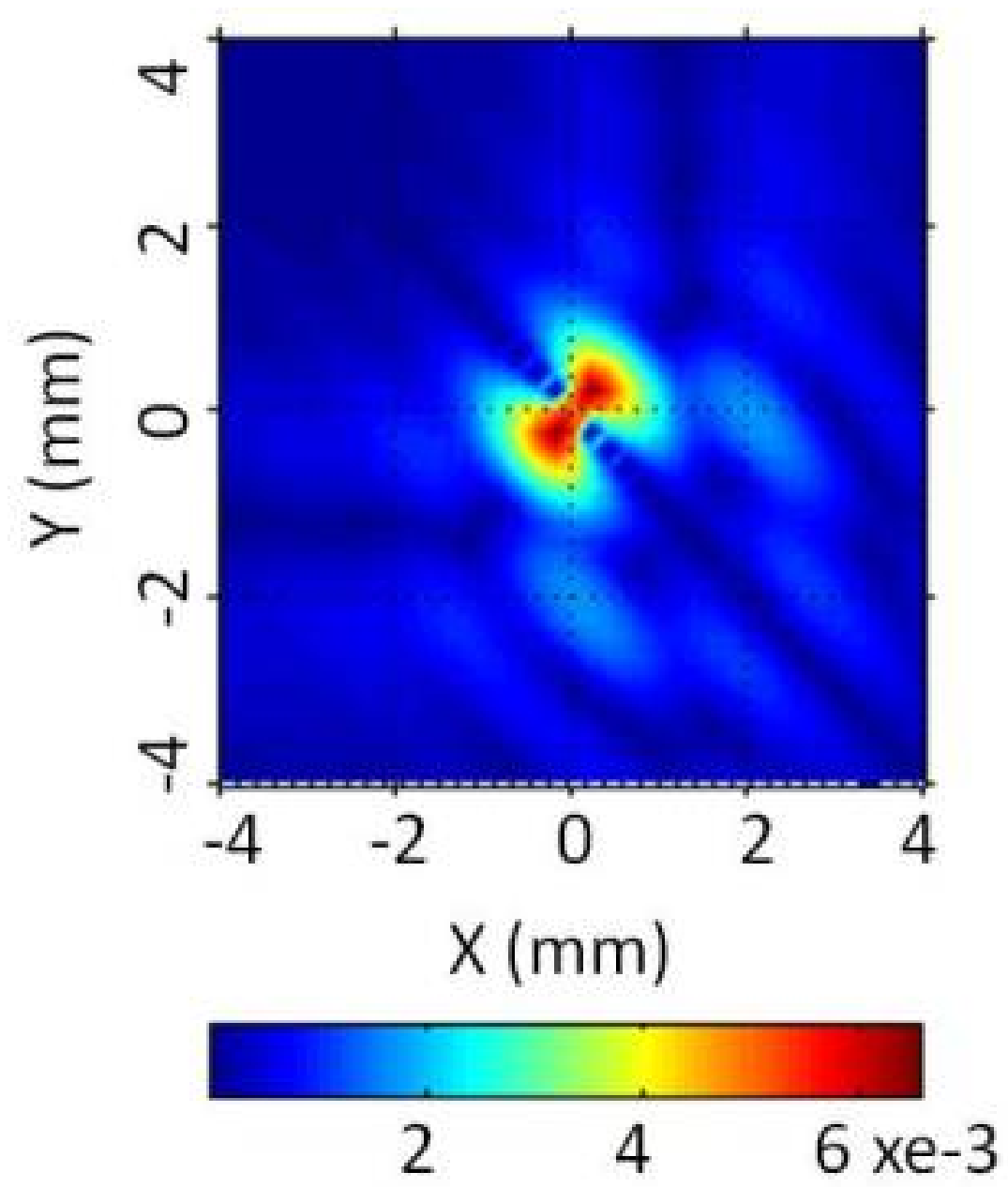}\\
(i) & (j) & (k) & (l)\\
\end{tabular}
\end{center}
\vspace{-0.1in}
\caption{Particle displacement per unit electrode area due to sectored annular transducers: (a)-(d) $u_z$: $90^{\circ}$, $120^{\circ}$, $180^{\circ}$, $270^{\circ}$; (e)-(h) $u_r$: $90^{\circ}$, $120^{\circ}$, $180^{\circ}$, $270^{\circ}$; (i)-(l)$u_\psi$: $90^{\circ}$, $120^{\circ}$, $180^{\circ}$, $270^{\circ}$.} 
\label{fig:UrPhiZSectoredFASAs}
\end{figure*}

We computed the particle displacements ($u_r$, $u_\psi$ and $u_z$) due to $90^{\circ}$, $120^{\circ}$,, $180^{\circ}$, and $270^{\circ}$ transducers. In all cases, the transducers are designed for focal length of 13mm at 3.85MHz and the maximum tile size of the transducer is limited to 8mm by 8mm; with these design parameters, the transducer can have only two-rings. It is worth emphasizing that the design parameters could be changed according to the user's requirements. All sectored transducers are placed around the origin (0,0), as shown in figure-\ref{fig:SectoredXducers}. The vertical particle displacements due to these transducers are shown in figure-\ref{fig:UrPhiZSectoredFASAs} (a), (b), (c), and (d); we notice that the $90^{\circ}$ transducer has two peaks, first peak near the origin and second peak at the edge of first ring. Also, we observe that the first peak has circular symmetry while the second maxima resembles the shape of the $90^{\circ}$ ring. Having distributed maxima of acoustic field or particle displacement helps in homogenous mixing of the fluid sample placed over the transducer. The vertical displacement component due to $120^{\circ}$, $180^{\circ}$, and $270^{\circ}$ are similar to $360^{\circ}$ and have only one prominent maxima at the origin. Such a localized acoustic field is useful in focused acoustic applications.

\begin{figure*}[t]
\begin{center}
\begin{tabular}{ccc}
\hspace{-0.1in}
\includegraphics[width=0.65\columnwidth]{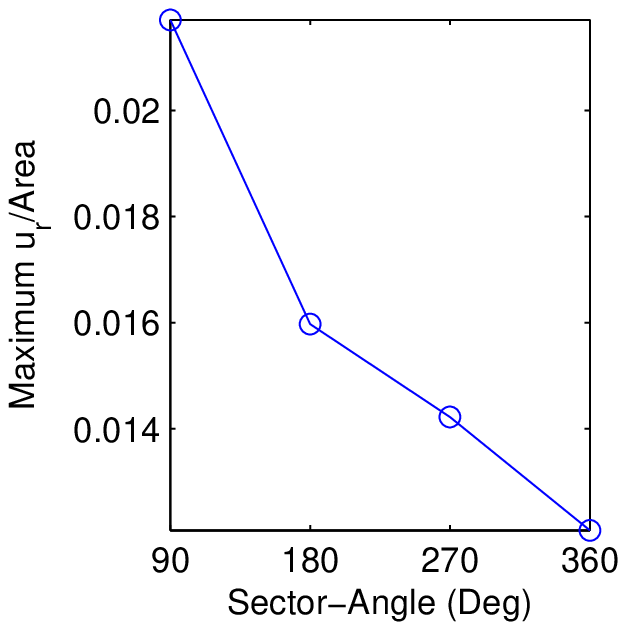} &
\includegraphics[width=0.65\columnwidth]{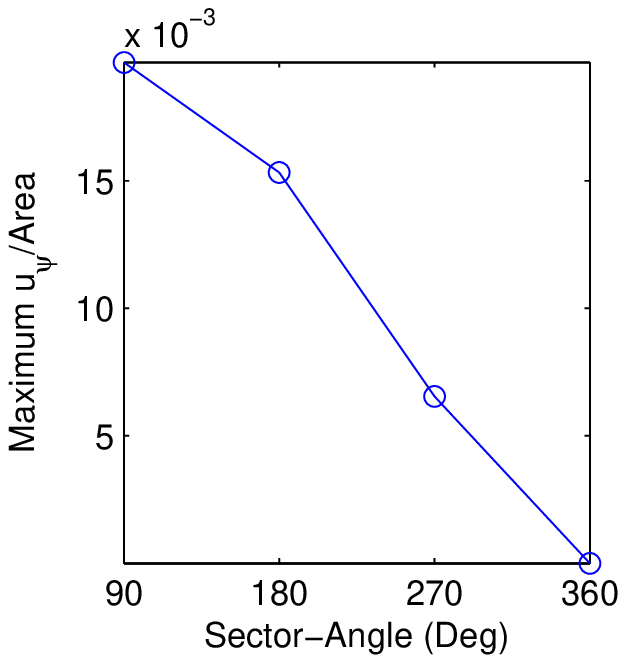} &
\includegraphics[width=0.65\columnwidth]{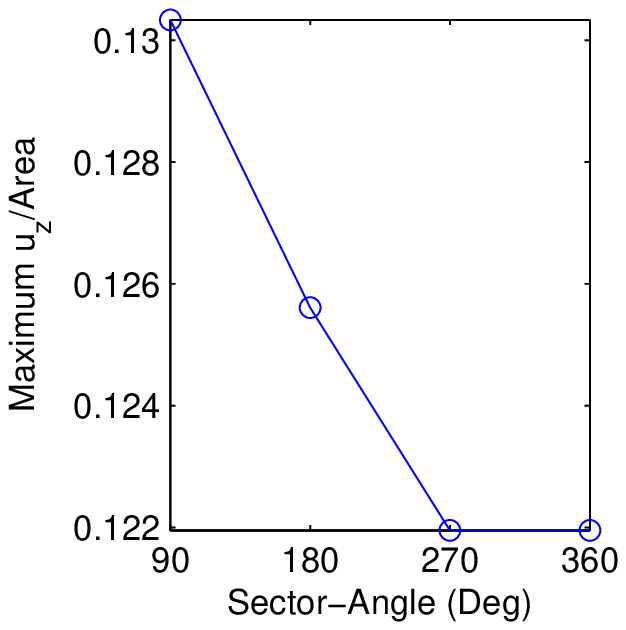}\\
(a) & (b) & (c)\\
\end{tabular}
\end{center}
\vspace{-0.1in}
\caption{Particle displacement per unit electrode area: (a) radial, $u_r$; (b) circumferential , $u_\psi$; (c) vertical component, $u_z$.} 
\label{fig:UrPhiZPerAreaSectoredFASAsPlot}
\end{figure*}

The radial component of particle displacements due to $90^{\circ}$, $120^{\circ}$, $180^{\circ}$, and $270^{\circ}$ sectored  transducers are shown in figure-\ref{fig:UrPhiZSectoredFASAs} (e), (f), (g), and (h). As expected, the peak of $u_r$ for a $90^{\circ}$ transducer, placed in the first quadrant, lies in the third quadrant because there is no active transducer element in the third quadrant to equalize the acoustic field generated towards third quadrant. Further, the $u_r$ due to $120^{\circ}$ has its maxima located mainly in first quadrant, but it is more distributed as compared to the $90^{\circ}$ one.  
Similarly, the $u_r$ due to $180^{\circ}$ has its maxima located in the two quadrants where the transducer is not having any active source of ultrasonic waves. As we increase the sector angle to more than $180^{\circ}$, the radial displacement have increased circular symmetry and therefore, $u_r$ of a $270^{\circ}$ transducer has its maximum intensity spread in three quadrants (I, III and IV in figure-\ref{fig:UrPhiZSectoredFASAs} (h)).

The circumferential component of particle displacement ($u_\psi$) due to sectored transducers has a unique characteristic. The maximum rotational component of a sectored transducer is always in the plane perpendicular to the orientation of the rings of the transducer. For example, the $90^{\circ}$ transducer placed in first quadrant has an orientation of $45^{\circ}$ and its $u_\psi$ has maximum intensity in $135^{\circ}$ and $315^{\circ}$ direction, as shown in figure-\ref{fig:UrPhiZSectoredFASAs} (i). Similarly, $u_\psi$ due to $120^{\circ}$, $180^{\circ}$ and $270^{\circ}$ transducers has peaks in the ($150^{\circ}$, $360^{\circ}$), ($0^{\circ}$, $180^{\circ}$) and ($45^{\circ}$, $225^{\circ}$) directions respectively. 

More importantly, all sectored transducers have higher rotational acoustic field than a full ring transducer. The maximum acoustic particle displacements ($u_r$, $u_\psi$, and $u_z$) normalized to electrode area is plotted in figure-\ref{fig:UrPhiZPerAreaSectoredFASAsPlot} and it is clear that the $90^{\circ}$ transducer has higher normalized field than $180^{\circ}$, $270^{\circ}$, and $360^{\circ}$ transducers. We also observe that the transducer with low sector-angle has higher normalized acoustic field; we select $90^{\circ}$ sector angle because we can build a  $180^{\circ}$, $270^{\circ}$, $360^{\circ}$ transducer by arranging more than one $90^{\circ}$ transducers in an array. For the same reasons, we did not choose the $120^{\circ}$ transducer as the basic transducer, as we could build only two sectored transducer from this-  $120^{\circ}$ and $240^{\circ}$. it could Also, we wanted to keep the number of transducers in an array to a low number to keep the requirement of number of separate RF signal low for the phased-excitation.

\section{Conclusion}
\label{sec:conclusion}
Efficient Design techniques for complex systems require accurate and efficient modeling for the components constituting the system. In this paper, we present an analytical modeling technique for piezoelectric transducers, which are major components in ultrasonic microfluid processing systems. We use this modeling approach for the design of a piston transducer which has both converging and vortexing properties. Namely, the proposed transducer electrodes are designed in the shapes of  rings cut into $90^{\circ}$, $120^{\circ}$, $180^{\circ}$ and $270^{\circ}$ sector angles. 

\vspace{-0.1in}
\bibliographystyle{IEEEtran}
\bibliography{paper_final}
\end{document}